\documentclass[12pt]{article}

\usepackage{amssymb,amsmath,amsthm,bbm,enumitem,xcolor,mathtools,dsfont}
\usepackage{graphicx,caption}
\usepackage{subcaption}
\usepackage[a4paper, total={6.3in, 9.2in}]{geometry}
\usepackage{setspace}
\usepackage{hyperref}
\usepackage{textcomp}
\usepackage[page]{appendix}

\usepackage{apacite} 
\AtBeginDocument{%

   \urlstyle{APACsame}%
}
\usepackage[round]{natbib}    

\newtheorem{theorem}{Theorem}[section]
\theoremstyle{definition}
\newtheorem{definition}{Definition}[section]
\theoremstyle{definition}

\theoremstyle{definition}
\newtheorem{assumption}{Assumption}[section]
\theoremstyle{definition}
\newtheorem{proposition}{Proposition}[section]
\theoremstyle{definition}

\theoremstyle{remark}

\theoremstyle{definition}
\newtheorem{lemma}{Lemma}[section]

\DeclareMathOperator{\real}{\mathbb{R}}

\DeclareMathOperator{\E}{\mathbb{E}}

\allowdisplaybreaks

\setlist[itemize]{leftmargin=0.4cm,labelindent=\parindent}

\title{Optimal Investment in Defined Contribution Pension Schemes with Forward Utility Preferences}
\author{Kenneth Tsz Hin Ng\thanks{Department of Mathematics, University of Illinois at Urbana-Champaign, Urbana, Illinois, United States. Email: \href{mailto:tszhinn2@illinois.edu}{tszhinn2@illinois.edu}.}    \;and   Wing Fung Chong\thanks{Maxwell Institute for Mathematical Sciences and Department of Actuarial Mathematics and Statistics, Heriot-Watt University, Edinburgh, Scotland, United Kindgom. Email: \href{mailto:alfred.chong@hw.ac.uk}{alfred.chong@hw.ac.uk}.}}
\date{\today}

\begin{document}
\maketitle

\begin{abstract}
Optimal investment strategies of an individual worker during the accumulation phase in the defined contribution pension scheme have been well studied in the literature. Most of them adopted the classical backward model and approach, but any pre-specifications of retirement time, preferences, and market environment models do not often hold in such a prolonged horizon of the pension scheme. Pre-commitment to ensure the time-consistency of an optimal investment strategy derived from the backward model and approach leads the supposedly optimal strategy to be sub-optimal in the actual realizations. This paper revisits the optimal investment problem for the worker during the accumulation phase in the defined contribution pension scheme, via the forward preferences, {\color{black}in which an environment-adapting strategy is able to hold optimality and time-consistency together.} 
Stochastic partial differential equation representation for the worker's forward preferences is illustrated. This paper  constructs two of the forward utility preferences and solves the corresponding optimal investment strategies, in the cases of initial power and exponential utility functions.
\end{abstract}




\section{Introduction}\label{sec:introduction}
Pension schemes are classified into two categories, namely the defined benefit (DB) and the defined contribution (DC), which compose the hybrid and the collective defined contribution schemes. On one hand, the DB scheme pre-defines a retirement benefit, while its plan members are required to contribute in accordance with the performance of the fund in order to maintain their future retirement benefits. On the other hand, the DC scheme asks its plan members to contribute a pre-determined proportion of their salaries to the fund, while their retirement benefits are based on, not only their portions of contribution from compensation, but also future investment returns by their customized portfolios. Shifts from the DB scheme to the DC scheme are evident in many countries. As the investment risk is fully borne by plan members in the DC scheme, seeking for the optimal investment strategies of its members has been a frontier research since the change of paradigm.\\

In the literature, optimal investment strategies of a worker (she/her) during her accumulation phase in the DC pension scheme were investigated with various objectives and quantities of interest. To name a few, \cite{BOULIER2001173}, \cite{HAN2012172:DC:inflation}, and \cite{GUAN201458:stochastic:interest:volatiltiy} maximized her expected utility on the difference between her pension fund value and an exogenous guarantee amount at the time of retirement. Instead of evaluating her pension fund value in absolute term, \cite{CAIRNS2006843:stochastic:lifestyle} suggested habit-formation in life-cycle models, that the worker would maximize her expected utility on her pension fund value with respect to her salary, as a num{\'e}raire, at her retirement time. Instead of assuming the classical expected utility as her preference, \cite{GUAN2016224:var:constraint} and \cite{CHEN2017137:minimum:performance} solved her optimal investment strategy by maximizing her expected $S$-shaped utility, on her absolute pension fund value in \cite{GUAN2016224:var:constraint}, and on the difference between her absolute fund value and an exogenous guarantee amount in \cite{CHEN2017137:minimum:performance}, both at her time of retirement. \cite{YAO201484:mean:variance} and \cite{GUAN201599:mean:variance:2} assumed the mean-variance objective for the worker, evaluating on her absolute pension fund value at her retirement time.\\

All of the aforementioned works, among others in the literature, adopted a classical backward model, and hence employed the classical backward approach to solve the optimal investment strategies. Any of these backward models sequentially pre-specify, at the current time when the optimal investment strategy is determined, (i) the worker's future time of retirement, (ii) her future preferences to evaluate her pension fund's performance, and (iii) models to govern the future dynamics between the current time and her retirement time. The worker's optimal investment strategy is then solved {\color{black}backward}, via, such as, dynamic programming principle (DPP), backward stochastic differential equation (BSDE) approach, or martingale method. Though it is solved {\color{black}backward}, her optimal investment strategy being planned is implemented {\color{black}forward} in time.\\

However, the revealed retirement time of the worker shall most likely be deviated from her assumption when the optimal investment strategy is planned. For example, in the United Kingdom, the State Pension age, which is the earliest age when a state pension could be claimed and acts as a proxy of retirement age, has been gradually increasing among generations. By the Pensions Act 2007, the State Pension age for women was increased from 60 to 65, over the period from April 2010 to April 2020. The Act also planned to increase the State Pension age for both men and women, from 65 to 66 phasing in between April 2024 and April 2026, from 66 to 67 between April 2034 and April 2036, and from 67 to 68 between April 2044 and April 2046. Also, her future preferences, assumed when the optimal investment strategy is planned, are most likely not able to capture the worker's change of risk appetite due to the actual realizations in the market environment. In addition, the assumed models for the market environment are certainly not going to imitate its true dynamics. For instance, it could be, microscopically, that a drift or a volatility of a dynamics, be it calibrated, does not behave as being revealed in real time; or, a systemic risk, such as pandemic or climate, is unfolded but was not taken into consideration; or, a state-of-the-art investment opportunity, such as green bond, and environmental, social, and governance (ESG) stock, emerges but was not available at the inception time of the worker's pension scheme.\\

These together drive the sub-optimality of the worker's planned investment strategy, which is supposed to be optimal. Given a long horizon, typically at least 40 years, this is particularly significant when determining investment strategies of pension plans. In order to {\color{black}forward} execute the {\color{black}backward} solved strategy, she must actually pre-commit the assumed backward model to ensure time-consistency of the strategy with respect to the model, but not necessarily to the actual realizations in the market environment. One might then incorporate updates on the backward model period-by-period, based on the latest experiences in the market environment, and, after each model-update, revise the investment strategy, which is again solved {\color{black}backward}, going forward. However, not only does such an approach violate the time-consistency of the strategies with respect to the models, as well as require substantial computational resources, but also, due to the classical backward approach to solve the strategies, an earlier portfolio adjustment action has already been contaminated by the mis-specified backward models of the market environment, albeit models being recurrently updated later. {\color{black}Section \ref{sec:backward:pitfall} in this paper shall briefly revisit the classical backward model and approach, and specifically highlight these pitfalls via a motivating example where there is a change of working environment being unfolded in the future, after the technical background of a stochastic control problem is set.}\\

Properly solving the optimal investment strategies during the accumulation phase of the DC pension scheme thus demands a call for forward model and approach. These should not pre-specify the worker's retirement time as well as her future preferences, and, while it is inevitable to assume a model for future dynamics, if model-update is regularly carried out, any portfolio adjustment actions before an update shall still be optimal. In other words, the optimal investment strategy should be time-consistent with respect to, not only the model assumed at the current time, but also the recurrently revised models, if any, using the actual realizations in the market environment. This would be feasible by a forward approach to solve the optimal investment strategy.\\

Forward preferences transpire to be a right notion for this. They were pioneered by \cite{Musiela2007,Musiela2008,Musiela2009,Musiela2010a,Musiela2010b,Musiela2011}, in which the forward preferences, as well as the corresponding optimal investment strategies, are constructed and solved for an agent (she/her) investing in a financial market, but without any considerations of her salary as well as its associated non-hedgeable risk which are crucial components during the accumulation phase of the DC pension scheme. First, the forward model pre-specifies, at the current time when the optimal investment strategy is determined, (i) the current preference of the agent, and (ii) a model of future dynamics in the market environment from the current time onward. The forward approach then, from the current time onward, constructs her future preferences, and hence the name as forward preferences, and solves her optimal investment strategy, via a forward induction on the Bellman optimality equation (which would be the backward induction on the equation, for DPP in the classical backward model and approach), or equivalently the forward version of the (super-)martingale (sub-)optimality principle. Therefore, in the forward model and approach, the agent's optimal investment strategy is solved and implemented {\color{black}forward} in the same direction of time. {\color{black} In addition, the optimal strategy at any time point depends only on the model assumption to-date, but not the model of the environment in the future. H}ence, any portfolio adjustment actions before a model-update, if any, are still optimal without any time-inconsistent issues, even though the agent does not fully commit to the assumed and recurrently updated, if any, models. \\

Since the above series of works by Musiela and Zariphopoulou, forward preferences have been extensively advanced in the literature on, for instance, dual characterization by \cite{zitkovi:2009}, stochastic partial differential equation (SPDE) representation by \cite{karoui:2013}, \cite{shkolnikov:2016}, and \cite{El:2018:FBSPDE}, homothetic processes by \cite{Nadtochiy2014}, ergodic BSDE and infinite horizon BSDE representations by \cite{liang:bsde}, model uncertainty using dual characterization by \cite{kallblad2018dynamically}, model uncertainty via direct construction by \cite{chong2019optimal}, mean-field games by \cite{lacker2019mean}, \cite{dosreisL:mean:field}, and \cite{dosreisL:mean:field2}, discrete-time binomial model by \cite{angoshtari:2020}, inverse investment problem by \cite{sigrid:inverse:investment}, rank-dependent preference by \cite{He:2021:rank:dependent}, Arrow-Pratt measure by \cite{strub2021evolution}, as well as investment and reinsurance by \cite{colaneri:katia:reinsurance:2021,colaneri:katia:reinsurance:2022}. Numerous applications of forward preferences can be observed in the literature on, maturity-independent risk measures by \cite{zitkovi:2010}, valuation of American options by \cite{leung2012forward}, indifference valuation in discrete-time binomial model by \cite{musiela2010indifference}, fund management by \cite{dynamic:preference:pension}, \cite{anthropelos:relative}, and \cite{hillairet:social}, forward entropic risk measures by \cite{chong2019ergodic}, pricing and hedging equity-linked life insurance by \cite{CHONG201993}, as well as robo-advising by \cite{robo:2021} and \cite{capponi2022personalized}.\\

This paper solves the optimal investment strategy of the worker enrolled to the DC pension scheme during the accumulation phase, using the forward utility preferences. Inspired by \cite{CAIRNS2006843:stochastic:lifestyle}, the worker evaluates her pension fund performance during the accumulation phase by the ratio of the fund value to her salary, ensuring that her pre-retirement habit could be consistent after any of her possible future retirement times. Given her current preference as her initial utility function on the ratio, her future utility preferences on the ratio are constructed, instead of being pre-specified, and her optimal investment strategy is correspondingly solved (see Definition \ref{def:forward} below). Again, by the forward model and approach, the planned investment strategy applies to any time moving forward, without the need of committing to any pre-specified retirement times; technically speaking, the solved optimal investment strategy is independent of any terminal times, and thus contributes the flexibility for the worker to decide her actual retirement time in due course. While one of the advantages for adopting the forward model and approach is, as mentioned above, the compatibility of model-update for future dynamics and time-consistency {\color{black}of environment-adapting strategy}, this paper solves the optimal investment strategy at the current time, with the focus on addressing the pre-specification issues, of (i) the worker's future retirement time and (ii) her future preferences, 
under the classical backward model and approach. {\color{black}To demonstrate the compatibility of model-updates using forward utility preferences, in Section \ref{sec:mot_eg_revisited}, we revisit the example in Section \ref{sec:backward:pitfall} and addresses the pitfalls of the backward approach. In general, i}ncorporating recurrent updates on the models for the market environment and regular revisions on the optimal time-consistent investment strategy using the forward preferences, as in \cite{robo:2021}, is very relevant to pension fund management, typically with a prolonged horizon; this shall be studied {\color{black}in details} as one of the future directions.\\

A long-standing technical challenge of the forward preferences is their non-uniqueness, and thus any associated problems are ill-posed; positively, this explains why individual preferences vary among agents. To illustrate the non-uniqueness of the worker's forward preferences and the corresponding optimal investment strategy, her (translated) forward preferences are characterized by an SPDE (see Equation \eqref{eq:SPDE} below), which particularly highlights the volatility processes of the (translated) forward preferences with respect to the financial market and the labour market. Unlike the endogenously implied volatility processes of the backward preferences derived from the backward model and approach, the volatility processes of the (translated) forward preferences are exogenously chosen by the worker herself at the current time, when she plans the optimal investment strategy going forward. A pair of volatility processes maps an SPDE for her forward preferences, which also then induces her optimal investment strategy (see Equation \eqref{eq:SPDE_strategy} below).\\

Following the SPDE representation, two forward preferences of the worker are constructed which solve the SPDE (see Equations \eqref{eq:U:power} and \eqref{eq:U:exp} below), and her corresponding optimal investment strategies are explicitly solved (see Equations \eqref{eq:pi*:power:1} and  \eqref{eq:pi*:exp} below). One case is when the worker's current preference on the ratio is given by the constant relative risk aversion (CRRA) power utility function, while the other case is specified by the constant absolute risk aversion (CARA) exponential utility function. In both cases, an exogenous baseline investment strategy emerges to drive the worker's forward preferences and her optimal strategy. Her forward preferences evaluate the comparison between, the ratio of her pension fund value to her salary, and the exogenous ratio generated by the baseline strategy. Such constructions are similar to relative performance criteria using the forward utility preferences in fund management of a financial market by \cite{anthropelos:relative}; see also \cite{dosreisL:mean:field} for the mean-field game setting. They constructed the forward utility preferences for one of the agents by comparing her fund performance to the other agent's (respectively, agents', in the mean-field game setting) performance, while the forward utility preferences of the worker constructed in this paper compare her fund performance to the exogenous performance, such as generated by a target-date fund. In the CRRA power utility case, any of the worker's admissible investment strategies leads to her pension fund value to salary ratio being bounded below by the exogenous ratio as a floor, which resembles the requirement in constructing the constant proportion portfolio insurance (CPPI) strategy in \cite{perold:1986:cppi}, \cite{black1987simplifying}, \cite{CPPI:1989}, and \cite{CPPI:1992}. In the CARA exponential utility case, the reciprocal of a similar exogenous ratio, also generated by the baseline strategy but without salary contribution, serves as a dynamic risk aversion process of the worker.  \\

As for the worker's optimal investment strategies, in both cases, they are in a form of (convex) combination of the baseline investment strategy and a myopic strategy, in which the myopic component resembles the optimal investment strategy in the respective cases under the classical backward model and approach. In the CRRA power utility case, her optimal investment strategy can also be viewed as a generalization of the CPPI strategy, with (i) the stochastic floor given by the exogenous baseline fund performance, (ii) a stochastic multiple of the cushion given by the myopic strategy, and (iii) an additive factor given by the amount of the exogenous baseline fund investing into a risky asset. Such a perspective does not hold for the CARA exponential utility case, as the other similar exogenous ratio, but without salary contribution, does not attribute as a floor of the worker's pension fund value to salary ratio. \\



To the best of knowledge, only \cite{dynamic:preference:pension} and \cite{hillairet:social} investigated pension management problems using the forward utility preferences. \cite{dynamic:preference:pension} constructed both CRRA power forward utility preferences, and symmetric asymptotic hyperbolic absolute risk aversion (SAHARA) forward utility preferences, of the worker, as well as solved her corresponding optimal CPPI and life-cycle investment strategies respectively. \cite{hillairet:social} proposed an investment-pension management problem for a social planner, who is endowed with forward utility preferences and solves the optimal investment-pension pair for her society, which consists of, workers who contribute their portion of compensation to a social pension system, and pensioners who receive their fixed amount of pension income from the system. This paper is different from \cite{dynamic:preference:pension} and \cite{hillairet:social}. Comparing to the former one, this paper takes the worker's salary contribution as well as its associated non-hedgeable risk into considerations, the worker herein evaluates her DC pension scheme fund performance by her fund value to salary ratio following the accumulation phase habit-formation advocate in \cite{CAIRNS2006843:stochastic:lifestyle}, and the worker's forward utility preference orders are caused by the comparison of her pension fund performance to the exogenous stochastic baseline. Comparing to the latter one, this paper solves an individual worker's pension fund management problem microscopically, while \cite{hillairet:social} concerns the social pension system at the macroscopic level to ensure sustainability and actuarial
fairness. \\

This paper is organized as follows. Section \ref{sec:problem} formulates the optimal investment problem of the worker enrolled to the DC pension scheme during the accumulation phase{\color{black}, and demonstrates the pitfalls in the classical backward modedl and approach.} {\color{black} Section \ref{sec:forward} revisits the forward preferences and} illustrates the SPDE representation of her (translated) forward preferences and highlights their volatility processes. Sections \ref{sec:power} and \ref{sec:exp} construct her forward utility preferences and solve her optimal investment strategies, in the cases of CRRA power and CARA exponential initial utility functions respectively. {\color{black}Section \ref{sec:new_section} recapitulates the reason to consider the worker's fund value to salary ratio as her quantity of interest, while also constructs the worker's forward utility preferences and optimal investment strategies based on the absolute pension fund value.} Section \ref{sec:conclusion} concludes and discusses future directions. {\color{black}For a clearer exposition of this paper, extensive derivations and proofs are relegated to Appendices \ref{sec:app:SPDE}, \ref{sec:app:proof:power}, and \ref{sec:app:proof:exp}.}



\section{Problem Formulation and {\color{black}Motivation}}
\label{sec:problem}

Let $(\Omega,\mathcal{F},\mathbb{P})$ be a probability space and ${\bf B}{\color{black}^1}=\{{\bf B}^1_t = (B^{1,j}_t)_{j=1}^n\}_{t\geq 0}$, ${\bf B}{\color{black}^2}=\{{\bf B}^2_t=(B^{2,j}_t)_{j=1}^m\}_{t\geq 0}$ be independent Brownian motions of dimensions $n$ and $m$, respectively.  Let $\mathbb{F}=\{\mathcal{F}_t\}_{t\geq 0}$ be the filtration generated by the Brownian motions $({\bf B}^1,{\bf B}^2)$ satisfying the usual conditions, and $\mathcal{P}_n(\mathbb{F})$ be the collection of all $\mathbb{F}$-progressively measurable processes taking values in $\mathbb{R}^n$. We also define
\begin{equation*}
\mathcal{L}^2_n := \Big\{ \pi=\{\pi_t = (\pi_t^i)_{i=1}^{n} \}_{t\geq 0} \in \mathcal{P}_n(\mathbb{F}):      \mathbb{E}\left[ \int_0^t \|\pi_s\|^2 ds   \right]<\infty, \text{ for all } t>0  \Big\},
\end{equation*}
where $\|\cdot\|$ is the Euclidean norm, and
\begin{equation*}
\mathcal{L}^2_{n,\text{BMO}} := \left\{ \pi\in \mathcal{P}_n(\mathbb{F}):\mathop{\text{esssup}}_{\tau\in\mathcal{T}[0,t]} \mathbb{E}\left[ \int_\tau^t \|\pi_s\|^2 ds \Big\vert \mathcal{F}_\tau   \right]<\infty, \text{ for all } t>0  \right\},
\end{equation*}
\begin{equation*}
\mathcal{T}[0,t]:=\left\{\tau\in \mathcal{P}_1(\mathbb{F}):\tau\text{ is an $\mathbb{F}$-stopping time in $\left[0,t\right]$}\right\},\text{ for all }t>0.
\end{equation*}


\subsection{Financial Market, Salary, and Pension Fund}\label{sec:market}
We consider a financial market with a riskless asset $S^0=\{S_t^0\}_{t\geq0}$, which earns a risk-free interest rate $r\in\mathbb{R}$, i.e., $S^0_t = S^0_0 e^{rt}$, and with $n$ risky assets ${\bf S} = \{{\bf S}_t = (S^i_t)_{i=1}^{n}\}_{t\geq 0}$, whose dynamics is given by, for any $i=1,2,\dots,n$, and $t\geq 0$,
\begin{equation*}
dS^i_t = S^i_t\left( (r  + \mu^i_t ) dt + \sum_{j=1}^n \sigma^{ij}_t dB^{1,j}_t\right). 
\end{equation*}
Herein, for any $i,j=1,2,\dots,n$, $\mu^i_{\cdot} : [0,\infty) \to \mathbb{R}$ represents the risk premium (instead of the mean in conventional notations) of the $i$-th risky asset, and $\sigma^{ij}_{\cdot} : [0,\infty) \to \mathbb{R}_+$ is the volatility of the $i$-th risky asset with respect to the $j$-th Brownian motion of ${\bf B}{\color{black}^1}$. For simplicity, both are assumed to be deterministic functions.\\

During the accumulation period of a pension fund, a worker receives salary, with a process $Y=\{Y_t\}_{t\geq 0}$ whose dynamics is given by, for any $t\geq 0$,
\begin{equation*}
dY_t = Y_t\left( (r+ \mu^Y_t)dt + (\sigma^{Y,1}_t)^\top d{\bf B}^1_t + (\sigma^{Y,2}_t)^\top d{\bf B}^2_t \right),\ Y_0=y>0,
\end{equation*}
where  $\mu^Y_{\cdot} :[0,\infty) \to \mathbb{R}$, $\sigma^{Y,1}_{\cdot} : [0,\infty) \to \mathbb{R}^{n}_+$, and $\sigma^{Y,2}_{\cdot} : [0,\infty) \to \mathbb{R}^{m}_+$ are, also for simplicity, assumed to be deterministic functions, which represent, respectively, the risk premium of the salary, and the volatility of the salary with respect to the Brownian motions ${\bf B}^1$ and ${\bf B}^2$. When $\sigma_{\cdot}^{Y,2}\not\equiv{\bf 0}$, the worker's salary is subject to risk(s) which could not be hedged using the risky assets, whence inducing market incompleteness. Despite its simplicity, this diffusion model serves as a reasonable approximation of a more realistic compound jump process for her salary dynamics. Such an approximation has also been applied in other actuarial contexts, such as in ruin theory; see, for instance, \cite{hanspeter1994diffusion,LUO2011123,COHEN2020333} and the references therein.\\

The worker is enrolled in a DC pension plan. Let $W=\{W_t\}_{t\geq0}$ be the process of her pension fund value. During the accumulation period and at any time $t\geq 0$, on one hand, the worker contributes part of her salary into the fund with an instantaneous rate $p_tY_t$, for some deterministic proportion of contribution $p_{\cdot}:\left[0,\infty\right)\rightarrow\mathbb{R}_+$; on the other hand, she self-manages her pension fund by actively investing in the financial market, in which $\pi_t^{i}$, for $i=1,2,\dots,n$, denotes the proportion of the fund investing into the $i$-th risky asset. 
The dynamics of $W$ is thus given by, for any $t\geq 0$,
\begin{equation}
dW_t = p_t Y_t dt  + W_t\left(   (r + \pi_t^\top \mu_t )dt + \pi^\top_t\Sigma_t d{\bf B}^1_t \right), \ W_0=w>0,
\label{eq:fund_value_dynamics}
\end{equation}
where $\pi_t = (\pi_t^i)_{i=1}^{n}$, $\mu_t = (\mu^i_t)_{i=1}^{n}$, and $\Sigma_t = ( \sigma^{ij}_t)_{i,j=1}^{n}$. The following assumptions are imposed in the remaining of this paper.
\begin{assumption}
\label{ass}
\quad
\begin{enumerate}
\item[(i)] The volatility matrix $\Sigma_t$ has a full rank for all $t\geq 0$;
\item[(ii)] The coefficients $\mu_{\cdot},\Sigma_{\cdot},\Sigma_{\cdot}^{-1}, \mu_{\cdot}^Y,\sigma_{\cdot}^{Y,1},\sigma_{\cdot}^{Y,2}$, the proportion of contribution $p_{\cdot}$, and the market price of risk $\lambda_{\cdot} := \Sigma_{\cdot}^{-1}\mu_{\cdot} $ are bounded in $t\geq 0$.
\end{enumerate}
\end{assumption}

In the context of retirement planning, the worker should perceive her living standard by the relative pension fund value to her pre-retirement salary, instead of by the absolute fund value{\color{black}. S}ee also \cite{CAIRNS2006843:stochastic:lifestyle}, particularly the life-cycle and habit-formation discussions therein{\color{black}; they suggest that, the absolute pension fund value should not be the sole factor when making a rational decision, while any changes of behaviour should be gradual rather than immediate. 
To this end, we assign the worker's salary as a  num\'{e}raire,} and define the pension fund value to salary ratio by $X=\{X_t := W_t/Y_t\}_{t\geq 0}$, which satisfies that, for any $t\geq 0$,
\begin{equation}
\begin{aligned}
dX_t =&\;p_tdt + X_t\left( \left( \pi^\top_t\Sigma_t (\lambda_t-\sigma^{Y,1}_t) - \mu^Y_t + \|\sigma^{Y,1}_t\|^2 + \|\sigma^{Y,2}_t\|^2 \right) dt  \right.\\&\quad\quad\quad\quad\quad\left.+ (\pi_t^\top\Sigma_t-(\sigma^{Y,1}_t)^\top) d{\bf B}^1_t - (\sigma^{Y,2}_t)^\top d{\bf B}^2_t \right),\\X_0=&\;x_0:= \frac{w}{y}>0.
\end{aligned}
\label{eq:X}
\end{equation}
Whenever necessary, we shall write $X^\pi=\{X^{\pi}_t\}_{t\geq 0}$ to emphasize its dependence on an investment strategy $\pi=\{\pi_t \}_{t\geq 0}$.\\


Let $\mathcal{A}$ be the admissible set of investment strategies in $\mathcal{P}_n(\mathbb{F})$, which shall minimally include an integrability condition, as well as a condition to ensure a preference of the worker being well-defined. For the former one, it shall be defined respectively in Sections \ref{sec:power_admissibility} and \ref{sec:admis_exponential}. For the latter one, in general, an admissible investment strategy $\pi$ should satisfy that
\begin{equation*}
X^\pi_t(\omega) \in \mathcal{D}_t(\omega),    \text{ for a.a. }(t,\omega)\in [0,\infty) \times\Omega,
\end{equation*}
where the stochastic domain $\mathcal{D} = \{\mathcal{D}_t\}_{t\geq 0}$ is an $\mathbb{F}$-progressively measurable set-valued process; in particular, the $\mathcal{F}_t$-measurable $\mathcal{D}_t\left(\cdot\right)$ is assumed throughout the paper to be an open random interval\footnote{Fix a time $t\geq 0$. The map $\omega \mapsto \mathcal{D}_t(\omega)$ is an  $\mathcal{F}_t$-measurable random open set in $\mathbb{R}$ if, for any compact set $F \subset \real$, it holds that  $  \{\omega \in\Omega : \left(\mathbb{R}\backslash \mathcal{D}_t(\omega)\right)\cap F \neq \varnothing\}\in \mathcal{F}_t$. Moreover, $\mathcal{D}_t\left(\cdot\right)$ is an open random interval, if $\mathcal{D}_t\left(\cdot\right)$ is a random open set and $\mathcal{D}_t(\omega)$ is an interval for almost all $\omega\in \Omega$. For details of the theory of random sets, see, such as, \cite{random:set}.} of the form $(\inf \mathcal{D}_t(\cdot),\infty)\subseteq\mathbb{R}$, for any time $t\geq 0$,  where $\{ \inf \mathcal{D}_t \}_{t\geq 0} \in \mathcal{P}_1(\mathbb{F})$. {\color{black}At the current time $t=0$, within the admissible set of strategies $\mathcal{A}$, the worker aims to seek the best one and plans to implement it {\color{black}forward}.}




{\color{black}\subsection{Motivating Example: Pitfalls of Classical Backward Model and Approach}
\label{sec:backward:pitfall}
As alluded in Section \ref{sec:introduction}, the classical backward model and approach dominate in the literature to solving the worker's optimal investment strategies; however, as a matter of fact, they suffer from either sub-optimality or time-inconsistent issue raised by her investment strategies, if there are any unexpected changes in the environment that are commonly observed in pension fund management problem with a prolonged horizon. This section briefly revisits the backward model and approach, and outlines a motivating example to illustrate these pitfalls.\\


At the current time $t=0$, the worker targets to retire at the future time $T>0$; she evaluates her future pension fund value to salary ratio by a pre-specified utility preference $u_T\left(\cdot\right)$; she assumes that the models for the financial market and her salary are governed as in Section \ref{sec:market} and pre-specifies all deterministic functions in these dynamics on the time horizon $\left[0,T\right]$. The worker then solves her optimal investment strategy from the following time-$0$ planning problem:
\begin{equation}
\sup_{\pi_{\left[0,T\right]}\in\mathcal{A}_{\left[0,T\right]}}\mathbb{E}\left[u_T\left(X^{\pi_{\left[0,T\right]}}_T\right)\right],
\label{eq:V:back:0}
\end{equation}
where $\pi_{\left[0,T\right]}$ and $\mathcal{A}_{\left[0,T\right]}$ are self-evident notations of an investment strategy and an admissible set on the time horizon $\left[0,T\right]$. For simplicity, suppose that the worker only invests in one risky asset (i.e., $n=1$, with denoting $\Sigma=\sigma$); she assumes that the deterministic functions in the dynamics are constant parameters, with $p_t=p>0$, and with $\sigma^{Y,2}=0$; her pre-specified utility preference is given by a CRRA power utility function (i.e., $u_T\left(x\right)=\frac{x^{\gamma}}{\gamma}$ for any $x\in\left(0,\infty\right)$ where $0<\gamma<1$, with $\mathcal{A}_{\left[0,T\right]}=\mathcal{L}^2_{1,\left[0,T\right]}$). By Case 3 in \cite{CAIRNS2006843:stochastic:lifestyle}, her optimal investment strategy $\pi_{\left[0,T\right]}^{\text{back}}=\{ \pi^{\text{back}}_{t,\left[0,T\right]}\}_{t\in[0,T]}$, which is solved by the backward approach, is given as, for any $t\in\left[0,T\right]$,
\begin{equation}
\label{eq:pi:back}
\pi^{\text{back}}_{t,\left[0,T\right]} := \frac{\sigma^{Y,1}}{\sigma}  + \frac{\lambda-\sigma^{Y,1}}{\sigma(1-\gamma)}  \left(1+\frac{pF_{t,T}\left(\mu^Y\right)}{X_t}\right), 
\end{equation}
where 
\begin{equation*}
F_{t,T}\left(\mu^Y\right) := \frac{e^{(\mu^Y-\lambda\sigma^{Y,1})(T-t)}-1}{\mu^Y-\lambda\sigma^{Y,1}}
\end{equation*}
is the unique risk-neutral price of a unit of her future salary receivable over the time horizon $[t,T]$. From the optimal solution \eqref{eq:pi:back} of the time-$0$ planning problem \eqref{eq:V:back:0}, the worker's optimal investment strategy depends on all of her pre-specifications at the current planning time $t=0$, which are her future retirement time $T$, her future utility preference via the CRRA parameter $\gamma$, and her assumed model dynamics on the time horizon $\left[0,T\right]$. Naturally, her investment strategy $\pi^{\text{back}}_{t,\left[0,T\right]}$ at any time $t\in\left[0,T\right]$, given in \eqref{eq:pi:back}, should depend on the assumed model dynamics on the time horizon $\left[0,t\right]$; yet, it depends on the assumed model dynamics over the future time horizon $\left[t,T\right]$ as well, particularly via her future salary receivable, which is an implicit consequence of the classical backward model and approach.\\

Suppose that the worker indeed {\color{black}forward} implemented the investment strategy $\pi_{\left[0,t_0\right)}^{\text{back}}$ given in \eqref{eq:pi:back} up to a time $t_0\in\left(0,T\right)$, and suppose further that, at this new current time $t=t_0$, the worker is promoted. This development results in a raise of her risk premium of the salary from $\mu^Y$ to $\Tilde{\mu}^Y$ applying over the future time horizon $\left[t_0,T\right]$, where $\Tilde{\mu}^Y>\mu^Y$, and ceteris paribus. Most importantly, the change is unexpected so it was not incorporated in the time-$0$ planning problem \eqref{eq:V:back:0}\footnote{\color{black}Promotion is just a practical instance. Technically, any unexpected changes in the parameters of the model dynamics would not have been incorporated in the time-$0$ planning problem \eqref{eq:V:back:0}.}. At the current time $t=t_0$, the worker could either:
\begin{enumerate}
\item[(i)] adapt to the changed environment, and {\color{black}forward} implement the investment strategy $\pi_{\left[t_0,T\right]}^{\text{back},1}=\{ \pi^{\text{back},1}_{t,\left[t_0,T\right]}\}_{t\in[t_0,T]}$ on the time horizon $\left[t_0,T\right]$ of the following time-$t_0$ planning problem, solved by the backward approach:
\begin{equation}
\sup_{\pi_{\left[t_0,T\right]}\in\mathcal{A}_{\left[t_0,T\right]}}\mathbb{E}_{t_0}\left[u_T\left(X^{\pi_{\left[t_0,T\right]}}_T\right)\right];
\label{eq:v:back:t0}
\end{equation}
her {\color{black}forward} implementing investment strategy on the time horizon $\left[0,T\right]$ would be given by $\pi_{\left[0,T\right]}^{\text{back},1}=\pi_{\left[0,t_0\right)}^{\text{back}}\oplus\pi_{\left[t_0,T\right]}^{\text{back},1}$, which is, for any $t\in\left[0,T\right]$,
\begin{equation}
\pi_{t,\left[0,T\right]}^{\text{back},1}:=
\begin{cases}
\frac{\sigma^{Y,1}}{\sigma}  + \frac{\lambda-\sigma^{Y,1}}{\sigma(1-\gamma)}  \left(1+\frac{pF_{t,T}\left(\mu^Y\right)}{X_t}\right),&\text{if }t\in\left[0,t_0\right);\\
\frac{\sigma^{Y,1}}{\sigma}  + \frac{\lambda-\sigma^{Y,1}}{\sigma(1-\gamma)}  \left(1+\frac{pF_{t,T}\left(\Tilde{\mu}^Y\right)}{X_t}\right),&\text{if }t\in\left[t_0,T\right];
\end{cases}
\label{eq:pi:back_1}
\end{equation}
\item[(ii)] commit to the pre-specified environment on solving the time-$0$ planning problem \eqref{eq:V:back:0}, and {\color{black}forward} implement the investment strategy $\pi_{\left[t_0,T\right]}^{\text{back},2}=\{ \pi^{\text{back},2}_{t,\left[t_0,T\right]}\}_{t\in[t_0,T]}$ on the time horizon $\left[t_0,T\right]$ as the same as the optimal investment strategy $\pi_{\left[t_0,T\right]}^{\text{back}}$ of the time-$0$ planning problem \eqref{eq:V:back:0}; her {\color{black}forward} implementing investment strategy on the time horizon $\left[0,T\right]$ would be given by $\pi_{\left[0,T\right]}^{\text{back},2}=\pi_{\left[0,t_0\right)}^{\text{back}}\oplus\pi_{\left[t_0,T\right]}^{\text{back}}=\pi_{\left[0,T\right]}^{\text{back}}$, which is given in \eqref{eq:pi:back}.
\end{enumerate}
Regardless of her choice, her investment strategy on the time horizon $\left[0,t_0\right)$ has already been implemented as $\pi_{\left[0,t_0\right)}^{\text{back},1}=\pi_{\left[0,t_0\right)}^{\text{back},2}=\pi_{\left[0,t_0\right)}^{\text{back}}$ given in \eqref{eq:pi:back}; the worker can no longer alter the decisions made, but could only regret them, if any.\\

On one hand, by definition, the investment strategy $\pi_{\left[t_0,T\right]}^{\text{back},2}$ is sub-optimal to the time-$t_0$ planning problem \eqref{eq:v:back:t0}, which is the latest concern of the worker at the current time $t=t_0$, given that the change of environment has unfolded. On the other hand, while, again by definition, the investment strategy $\pi_{\left[t_0,T\right]}^{\text{back},1}$ is optimal to the time-$t_0$ planning problem \eqref{eq:v:back:t0}, it leads to a time-inconsistent issue, with the reasons as follows.\\

First of all, by \eqref{eq:pi:back} and \eqref{eq:pi:back_1}, it is obvious that $\pi_{\left[t_0,T\right]}^{\text{back},1}\neq\pi_{\left[t_0,T\right]}^{\text{back}}$. Second, if the worker had oracularly anticipated, at the past time $t=0$, her promotion and hence the raise of the parameter value of the risk premium of her salary from $\mu^Y$ to $\Tilde{\mu}^Y$ at the current time $t=t_0$, her optimal investment strategy $\pi_{\left[0,T\right]}^{\text{back},*}=\{ \pi^{\text{back},*}_{t,\left[0,T\right]}\}_{t\in[0,T]}$ of the time-$0$ planning problem \eqref{eq:V:back:0}, solved by the backward approach, would have been given as, for any $t\in\left[0,T\right]$,
\begin{equation}
\label{eq:pi:back_star}
\pi^{\text{back},*}_{t,\left[0,T\right]} := \frac{\sigma^{Y,1}}{\sigma}  + \frac{\lambda-\sigma^{Y,1}}{\sigma(1-\gamma)}  \left(1+\frac{p\tilde{F}_{t,T}\left(\mu^Y,\Tilde{\mu}^Y\right)}{X_t}\right), 
\end{equation}
where 
\begin{equation*}
\tilde{F}_{t,T}\left(\mu^Y,\Tilde{\mu}^Y\right):=
\begin{cases}
F_{t,t_0}\left(\mu^Y\right)+e^{\left(\mu^Y-\lambda\sigma^{Y,1}\right)\left(t_0-t\right)}F_{t_0,T}\left(\Tilde{\mu}^Y\right),&\text{if }t\in\left[0,t_0\right);\\
F_{t,T}\left(\Tilde{\mu}^Y\right),&\text{if }t\in\left[t_0,T\right].
\end{cases}
\end{equation*}
Therefore, by \eqref{eq:pi:back}, \eqref{eq:pi:back_1}, and \eqref{eq:pi:back_star}, not only $\pi_{\left[t_0,T\right]}^{\text{back},1}=\pi_{\left[t_0,T\right]}^{\text{back},*}\neq\pi_{\left[t_0,T\right]}^{\text{back}}$, but also $\pi_{\left[0,t_0\right)}^{\text{back},*}\neq\pi_{\left[0,t_0\right)}^{\text{back}}$. In fact, the latter inequality is a consequence of the former inequality, due to the backward approach on solving the time-$0$ planning problem \eqref{eq:V:back:0}, whether or not the worker is an oracle; indeed, since $\pi_{\left[t_0,T\right]}^{\text{back},*}\neq\pi_{\left[t_0,T\right]}^{\text{back}}$, their {\color{black}backward} induced value functions at the time $t=t_0$, of the time-$0$ planning problem \eqref{eq:V:back:0}, are not the same{\color{black}. As} a result, $\pi_{\left[0,t_0\right)}^{\text{back},*}$ and $\pi_{\left[0,t_0\right)}^{\text{back}}$ are different, even though the environments are the same on the time horizon $\left[0,t_0\right]$. This latter inequality sheds light on the time-inconsistency that, at the current time $t=t_0$ with the new development being revealed in the environment, the worker regrets her implemented investment strategy $\pi_{\left[0,t_0\right)}^{\text{back}}$ given in \eqref{eq:pi:back} on the time horizon $\left[0,t_0\right)$; if she had prophesied the exact change in the environment, she could have implemented the investment strategy $\pi_{\left[0,t_0\right)}^{\text{back},*}$ given in \eqref{eq:pi:back_star} on the time horizon $\left[0,t_0\right)$, rather than $\pi_{\left[0,t_0\right)}^{\text{back}}$. This time-inconsistent issue stems from the backward approach, together with fact that $\pi_{\left[t_0,T\right]}^{\text{back},1}\neq\pi_{\left[t_0,T\right]}^{\text{back}}$; indeed, if $\pi_{\left[t_0,T\right]}^{\text{back},1}=\pi_{\left[t_0,T\right]}^{\text{back}}=\pi_{\left[t_0,T\right]}^{\text{back},*}$, since the environments are the same on the time horizon $\left[0,t_0\right]$, by the backward approach on solving the time-$0$ planning problem \eqref{eq:V:back:0} via their same {\color{black}backward} induced value functions at the time $t=t_0$, $\pi_{\left[0,t_0\right)}^{\text{back},*}$ and $\pi_{\left[0,t_0\right)}^{\text{back}}$ would be equal to each other; that is, the worker would not regret her implemented investment strategy. Yet, $\pi_{\left[t_0,T\right]}^{\text{back},1}=\pi_{\left[t_0,T\right]}^{\text{back}}$ implies that her investment strategy on the time horizon $\left[t_0,T\right]$ is the pre-commitment strategy $\pi_{\left[t_0,T\right]}^{\text{back},2}$, which has been discussed to be sub-optimal to the time-$t_0$ planning problem \eqref{eq:v:back:t0}.\\

These illustrate that the classical backward model and approach suffer from either the sub-optimality of the pre-commitment strategy $\pi_{\left[t_0,T\right]}^{\text{back},2}$ or the time-inconsistent issue raised by the environment-adapting strategy $\pi_{\left[t_0,T\right]}^{\text{back},1}$.\\

To further demonstrate the time-inconsistent issue induced by the environment-adapting strategy, the remaining of this motivating example studies the difference $\pi_{t,\left[0,T\right]}^{\text{back},*}-\pi_{t,\left[0,T\right]}^{\text{back}}$, for $t\in\left[0,t_0\right)$, when they are {\color{black}forward} implemented by the worker. Suppose that $r=3\%$, $\mu^{1}_{\cdot}=8\%$, and $\Sigma_{\cdot}=\sigma_{\cdot}=20\%$ (and thus, $\lambda_{\cdot}=0.4$); the values of the parameters in the financial market model are largely comparable to those being used in the literature on pension fund management; see, for example, \cite{CAIRNS2006843:stochastic:lifestyle}, \cite{BARUCCI20125588}, and the references therein. The worker targets to retire at time $T=20$. Her initial fund value to salary ratio is set as $x_0=1$, while she contributes a constant proportion\footnote{\color{black}The Fidelity\textsuperscript{\textregistered} Q1 2023 Retirement Analysis indicates that, the average contribution rate for 401(k) plan in the United States is 14\% for the first quarter of 2023, which is slightly higher than the figure $13.7\%$ for the fourth quarter of 2022.} $p_{\cdot}=10\%$ throughout the accumulation period $\left[0,20\right]$. Suppose further that $\mu^Y_\cdot =2\%$ (and thus the growth rate of her salary is given by $r+\mu^Y_\cdot=5\%$) and $\sigma^{Y,1}_{\cdot}=8\%$, for the values of the parameters in the salary model; the former aligns with the Employment Cost Index published by the United States Bureau of Labor Statistics\footnote{\color{black}According to the United States Bureau of Labor Statistics (see, for instance, \href{https://www.bls.gov/eci/latest-numbers.htm}{https://www.bls.gov/eci/latest-numbers.htm}), the 12-month wage increases for the first quarter of 2023 is $4.8\%$ for both private industry and civilian workers.}, while the latter volatility with respect to the hedgeable risk is set to be lower than $10\%$, which assumes that her salary is less volatile than the risky asset in the financial market. The worker's CRRA parameter $\gamma=0.6$. Assume that her promotion will take place at time $t_0=10$, when the risk premium of her salary will be increased from $\mu^Y_\cdot =2\%$ to $\tilde{\mu}^Y_\cdot = 7\%$, applying to the remaining accumulation period $\left[10,20\right]$; bear in mind that the worker would not have known this change of environment at time $t=0$. To study the difference $\pi_{t,\left[0,20\right]}^{\text{back},*}-\pi_{t,\left[0,20\right]}^{\text{back}}$, for $t\in\left[0,10\right)$, $10,000$ samples are drawn, and Figure \ref{fig:CDFs:backward} depicts its empirical cumulative distribution function (CDF) at time $t=5$ and $t=9$ based on these $10,000$ samples. From Figure \ref{fig:CDF:5}, there are $9,822$ samples, out of the $10,000$ samples, that the difference is positive at time $t=5$; this implies that, if the worker had expected that the risk premium of her salary is going to be increased from $\mu^Y_\cdot =2\%$ to $\tilde{\mu}^Y_\cdot = 7\%$ at time $t_0=10$, then, with a very high probability of $0.9822$, she would have invested more on the risky asset. From Figure \ref{fig:CDF:9}, at the later time $t=9$ being closer to the time when the risk premium of her salary increases, the probability, that the worker would have invested more on the risky asset if she had anticipated the change of environment, is only mildly reduced to $0.9745$; this suggests that the time-inconsistent issue remains to be significant throughout the time horizon $\left[0,10\right)$ before the new development is unfolded.\\

\begin{figure}[!h]
\centering
\begin{subfigure}{.5\textwidth}
\centering
\includegraphics[scale=0.4]{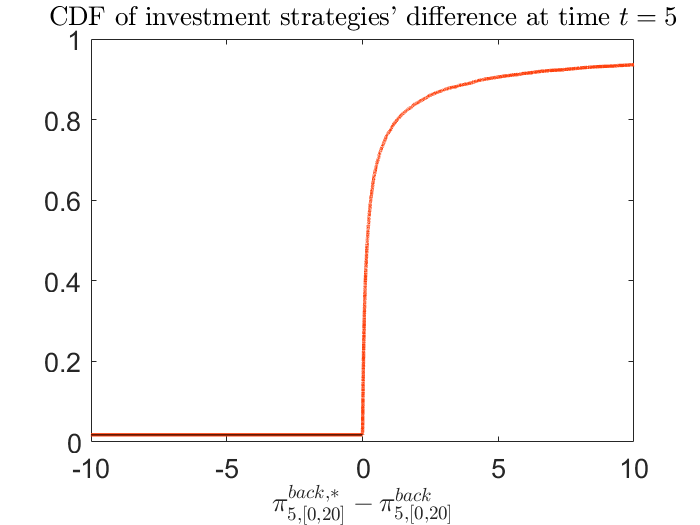}
\caption{\color{black}$t=5$}
\label{fig:CDF:5}
\end{subfigure}%
\begin{subfigure}{.5\textwidth}
\centering
\includegraphics[scale=0.4]{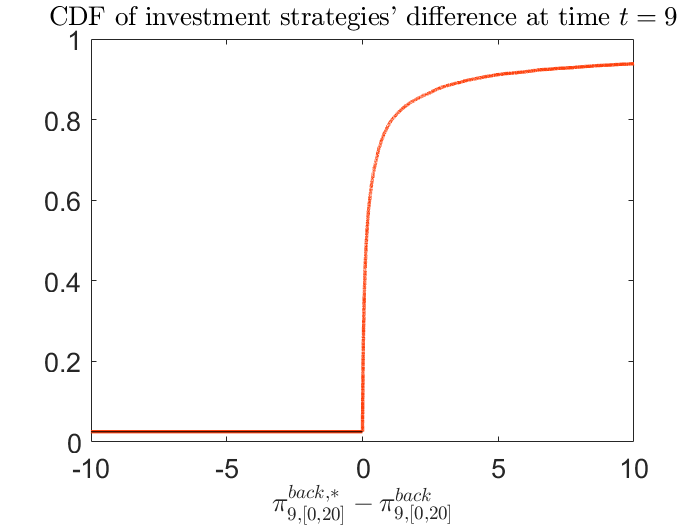}
\caption{\color{black}$t=9$}
\label{fig:CDF:9}
\end{subfigure}%
\caption{\color{black}Empirical cumulative distribution functions of investment strategies' difference $\pi_{t,\left[0,20\right]}^{\text{back},*}-\pi_{t,\left[0,20\right]}^{\text{back}}$, at times $t=5$ and $t=9$, before the change of environment}
\label{fig:CDFs:backward}
\end{figure}

This motivating example shall be revisited in Section \ref{sec:mot_eg_revisited} under the forward model and using the forward approach.}

 \section{Forward Preferences}
\label{sec:forward}
{\color{black} With the pitfalls of using a backward utility preference demonstrated in Section \ref{sec:backward:pitfall}, we introduce the use of forward preferences in managing a DC pension plan to address the aforementioned issues, which shall be discussed in details in Section \ref{sec:mot_eg_revisited}. Specifically, t}he worker aims to solve an optimal investment strategy, using her forward preference on the fund value to salary ratio $X$ as the optimality criterion. The notion of forward preference is essentially a stochastic process satisfying a martingale (resp. super-martingale) property when it evaluates an optimal (resp. a sub-optimal) fund value to salary ratio. The preference models the worker's future utilities on the ratio moving forward in the accumulation period, which is subject to both financial market and salary performances and thus explains its nature of stochasticity. The following definition recalls such notion, with a slight modification from the original definition in \cite{Musiela2007} to incorporate the restriction for the fund value to salary ratio lying in $\mathcal{D}$.


\begin{definition}
\label{def:forward}
Let $\mathcal{D}=\{\mathcal{D}_t\}_{t\geq 0}$ be an open set-valued process  with $\mathcal{D}_t=(\inf \mathcal{D}_t,\infty)$ for any $t\geq 0$, where $\{ \inf \mathcal{D}_t \}_{t\geq 0} \in \mathcal{P}_1(\mathbb{F})$. A random field $U=\{U(x,t;\omega) :  \omega\in\Omega, \ t \geq 0,\ x\in  \mathbb{R} \}$ is called a forward preference on the fund value to salary ratio defined in  $\mathcal{D}$, if  it satisfies all of the following properties:
\begin{enumerate}[label=(\roman*)]
\item for each $x\in\mathbb{R}$, $U(x,\cdot;\cdot)$ is $\mathbb{F}$-adapted;
\item for each $t \geq 0$ and $\omega\in\Omega$, $x \in \mathcal{D}_t(\omega) \mapsto U(x,t;\omega)$ is strictly increasing and strictly concave, while, for $x\leq   \inf \mathcal{D}_t(\omega)$, $U(x,t;\omega) = -\infty$.
\item for any $\pi \in \mathcal{A}$, and $0\leq s\leq t$,
\begin{equation*}
U(X_s^\pi,s) \geq \E[U(X_t^\pi,t) \vert\mathcal{F}_s], \  \mathbb{P}\text{-a.s.},
\end{equation*}
and there exists an optimal investment strategy $\pi^* \in \mathcal{A}$ such that, for any $0\leq s\leq t$, 
\begin{equation*}
U(X_s^{\pi^*},s) = \E[U(X_t^{\pi^*},t) |\mathcal{F}_s], \  \mathbb{P}\text{-a.s.}.
\end{equation*}
\end{enumerate}
\end{definition} 

\subsection{Volatility Processes of Forward Preferences} 
\label{sec:SPDE}
The stochastic nature of the worker's forward preference on her fund value to salary ratio $X$ can be revealed by a SPDE representation. Let $U$ be a random field satisfying (i) and (ii) in Definition \ref{def:forward} with stochastic domain $\mathcal{D}$. Consider the following two cases:
    \begin{enumerate}
        \item[(i)]  if $\inf \mathcal{D}_0 > -\infty$, $\mathbb{P}$-a.s., let the process $Z:=\{Z_t:= \inf \mathcal{D}_t\}_{t\geq 0}$ which admits the It\^o's diffusion form: for any $t\geq 0$,
         \begin{equation}
         \label{eq:Z:SPDE}
        dZ_t = \nu_t dt + (\kappa^1_t)^\top d{\bf B}^1_t + (\kappa^2_t)^\top d{\bf B}^2_t.
    \end{equation}
    where $\nu=\{\nu_t \}_{t\geq 0}\in \mathcal{P}_1(\mathbb{F})$, $\kappa^1=\{\kappa^1_t \}_{t\geq 0}\in \mathcal{P}_n(\mathbb{F})$ and $\kappa^2=\{\kappa^2_t \}_{t\geq 0}\in \mathcal{P}_m(\mathbb{F})$, and let $\mathcal{\tilde{D}} := \mathbb{R}_{++}$, which is $\mathbb{R}_+\backslash\{0\}$.
    \item[(ii)] if $\inf \mathcal{D}_0 = -\infty$, $\mathbb{P}$-a.s., let $\mathcal{D} \equiv \mathbb{R}$ and the process $Z:\equiv 0$ (which is equivalent to $Z$ satisfying \eqref{eq:Z:SPDE} with $Z_0=0$ and coefficients $\nu$, $\kappa^1$ and $\kappa^2$ all being identical to zero), and let $\mathcal{\tilde{D}} := \mathbb{R}$.
    \end{enumerate}
To unify the subsequent discussions, define the translated random field $\tilde{U}=\{\tilde{U}(\tilde{x},t;\omega) : \omega \in\Omega, \ t\geq 0, \ \tilde{x} \in \mathcal{\tilde{D}}\}$ by $\tilde{U}(\tilde{x},t):= U(\tilde{x}+Z_t,t)$, which obviously satisfies (i) and (ii) in an equivalent form of Definition \ref{def:forward}. Notice that  $U$ is a forward preference on $X$ with stochastic domain $\mathcal{D}$ if and only if $\tilde{U}$ is a forward preference on $\tilde{X}:=\{\tilde{X}_t:=X_t-Z_t\}_{t\geq 0}$ with deterministic domain $\mathcal{\tilde{D}}$.  Assume that $\tilde{U}$ takes the following It\^o's diffusion form: for any $\tilde{x}\in\tilde{\mathcal{D}}$ and $t\geq 0$,
\begin{equation*}
d\tilde{U}(\tilde{x},t) =  b(\tilde{x},t) dt + a_1(\tilde{x},t)^\top d{\bf B}^1_t + a_2(\tilde{x},t)^\top d{\bf B}^2_t, \ 
\tilde{U}(\tilde{x},0) =  u_0(\tilde{x}+Z_0), 
\label{eq:U}
\end{equation*}
where, for each $\tilde{x}\in\tilde{\mathcal{D}}$, $b\left(\tilde{x},\cdot;\cdot\right)\in\mathcal{P}_1(\mathbb{F})$, $a_1\left(\tilde{x},\cdot;\cdot\right)\in\mathcal{P}_n(\mathbb{F})$, and $a_2\left(\tilde{x},\cdot;\cdot\right)\in\mathcal{P}_m(\mathbb{F})$, and where  $u_0\left(\cdot\right)$ is a deterministic function mapping from $\mathcal{D}_0$ to $\mathbb{R}$. Suppose further that, for each $t\geq 0$ and $\omega\in\Omega$, $a_1\left(\cdot,t;\omega\right)$ and $a_2\left(\cdot,t;\omega\right)$ are differentiable in $\mathcal{\tilde{D}}$. \\

At the current time $0$, the worker's utility on the ratio is given by the deterministic $u_0\left(\cdot\right)$. The random fields $a_1=\left\{a_1\left(\tilde{x},t;\omega\right)\right\}_{\omega\in\Omega,\tilde{x}\in\mathcal{\tilde{D}},t\geq 0}$ and $a_2=\left\{a_2\left(\tilde{x},t;\omega\right)\right\}_{\omega\in\Omega,\tilde{x}\in\mathcal{\tilde{D}},t\geq 0}$ represent the volatility processes of the random field $\tilde{U}$ with respect to the hedgeable component of the financial market and the non-hedgeable component of the labour market respectively, which are chosen by the worker at the current time $0$. The random field $b=\left\{b\left(\tilde{x},t;\omega\right)\right\}_{\omega\in\Omega,\tilde{x}\in\mathcal{\tilde{D}},t\geq 0}$, which is the drift process of the random field $\tilde{U}$, is then determined to ensure that the equivalent form of Definition \ref{def:forward}, particularly (iii), is satisfied moving forward, and hence $\tilde{U}$ is a forward preference. With proper transformation, $U$ is then a forward preference.\\


{\color{black}By the It\^o-Wentzell formula, with the detailed derivations being relegated to Appendix \ref{sec:app:SPDE},} the translated forward preference $\tilde{U}$ satisfies the SPDE: for any $\tilde{x}\in\mathcal{\tilde{D}}$ and $t\geq 0$,
\begin{equation}
\begin{aligned}
d\tilde{U}(\tilde{x},t) =&\; 
\Bigg(\left((\tilde{x}+Z_t)\sigma^{Y,2}_t+ \kappa_t^2 \right)^\top \nabla_{\tilde{x}}a_2(\tilde{x},t) - \frac{ \|(\tilde{x}+Z_t)\sigma^{Y,2}_t+\kappa_t^2\|^2\tilde{U}_{\tilde{x}\tilde{x}}(\tilde{x},t)}{2} \\ & \ \quad -  \tilde{U}_{\tilde{x}}(\tilde{x},t) (p_t-\nu_t + (\lambda_t-\sigma^{Y,1}_t)^\top \kappa^1_t+ (\tilde{x}+Z_t)(\lambda_t^\top\sigma^{Y,1}_t+\|\sigma^{Y,2}_t\|^2 - \mu^Y_t))    \\ &\ \quad  + \frac{ \left\|\nabla_{\tilde{x}} a_1(\tilde{x},t)+ (\lambda_t-\sigma^{Y,1}_t)\tilde{U}_{\tilde{x}}(\tilde{x},t)   \right\|^2   }{2\tilde{U}_{\tilde{x}\tilde{x}}(\tilde{x},t)}  \Bigg) dt \\ &\; + a_1^\top(\tilde{x},t) d{\bf B}^1_t + a^\top_2(\tilde{x},t) d{\bf B}^2_t,
\end{aligned} 
\label{eq:SPDE}
\end{equation}
and the forward preference $U$ is given by, for any $t\geq 0$,
    \begin{equation}
        U(x,t) = \begin{cases}
            \Tilde{U}(x-Z_t,t), &\text{if }x\in\mathcal{D}_t;\\
            -\infty, &\text{otherwise}.
        \end{cases}
        \label{eq:translation_U}
    \end{equation}
Furthermore, {\color{black}the optimal investment strategy $\pi^*$ is given by}, for any $t\geq 0$,
\begin{equation}
\begin{aligned}
\pi_t^* :=&\; (\Sigma^\top_t)^{-1}\left(\sigma^{Y,1}_t-\frac{\nabla_{\tilde{x}} a_1(\tilde{X}_t,t)+ (\lambda_t-\sigma^{Y,1}_t)\tilde{U}_{\tilde{x}}(\tilde{X}_t,t) -\kappa^1_t \tilde{U}_{\tilde{x}\tilde{x}}(\tilde{X}_t,t) }{ X_t \tilde{U}_{\tilde{x}\tilde{x}}(\tilde{X}_t,t) }\right)
\\=&\;  (\Sigma^\top_t)^{-1} \left(\sigma^{Y,1}_t- \frac{\nabla_{\tilde{x}} a_1(X^{\pi^*}_t-Z_t,t)+ (\lambda_t-\sigma^{Y,1}_t)U_x(X^{\pi^*}_t,t)-\kappa^1_t U_{xx}(X^{\pi^*}_t,t)}{ X^{\pi^*}_t U_{xx}(X^{\pi^*}_t,t) }\right).
\end{aligned}
\label{eq:SPDE_strategy}
\end{equation}
{\color{black}For the existence of a unique solution for the SPDE in Equation \eqref{eq:SPDE}, we refer interested readers to \cite{karoui:2013} and \cite{El:2018:FBSPDE}.}\\

After motivating the volatility processes of forward preferences, we shall construct forward preferences by incorporating exogenous stochastic input factors into an \textit{ansatz} and identify their corresponding volatility processes $a_1$ and $a_2$. The following worker's initial utility are considered respectively in Sections \ref{sec:power} and \ref{sec:exp}.
\begin{itemize}
\item Power utility:
\begin{equation*}
u_0\left(x\right)=
\begin{cases}
\frac{1}{\gamma}x^{\gamma} & \text{for }x\in\mathcal{D}_0\\
-\infty & \text{for }x\in\mathbb{R}\backslash\mathcal{D}_0\\
\end{cases}
\end{equation*}
where $0<\gamma<1$ and $\mathcal{D}_0=\left(0,\infty\right)$.
\item Exponential utility:
\begin{equation*}
u_0\left(x\right)=-\exp\left(-\gamma x\right)\quad\text{for }x\in\mathcal{D}_0=\mathbb{R},
\end{equation*}
where $\gamma>0$.
\end{itemize}

\section{Power Forward Utility Preferences} 
\label{sec:power}
To construct power forward utility preference, the stochastic domain $\mathcal{D}$ is bounded from below, with $\inf\mathcal{D}_0=0$. Let $Z=\{Z_t:= \inf \mathcal{D}_t\}_{t\geq 0}$ be the greatest lower bound of the domain $\mathcal{D}$. {\color{black}To motivate the specific It{\^o}'s diffusion form that $Z$ should satisfy in \eqref{eq:Z:SPDE}, observe that the transformation in \eqref{eq:translation_U} suggests that the greatest lower bound process $Z$ resembles an exogenous minimum guarantee, which has been modelled in the literature to be surpassed by the performance of the worker's investment strategy leading to her utility; see, for example, \cite{BOULIER2001173}, \cite{HAN2012172:DC:inflation}, \cite{GUAN201458:stochastic:interest:volatiltiy}, and \cite{CHEN2017137:minimum:performance} in the backward model and approach.\\

To this end, let $\hat{\pi}=\left\{\hat{\pi}_t\right\}_{t\geq 0}\in \mathcal{P}_n\left(\mathbb{F}\right)$ be an exogenously chosen baseline investment strategy, which is assumed to be uniformly bounded. In turn, the greatest lower bound process $Z$, which is the minimum guarantee of a fund value to salary ratio under the baseline strategy, should thus follow the same dynamics as $X$ by implementing the strategy $\hat{\pi}$; that is, $Z_0=0$, and, for any $t\geq 0$,
\begin{equation}
\begin{aligned}
dZ_t =&\;p_tdt + Z_t\left( \left( \hat{\pi}^\top_t\Sigma_t (\lambda_t-\sigma^{Y,1}_t) - \mu^Y_t + \|\sigma^{Y,1}_t\|^2 + \|\sigma^{Y,2}_t\|^2 \right) dt  \right.\\&\quad\quad\quad\quad\quad\left.+ (\hat{\pi}_t^\top\Sigma_t-(\sigma^{Y,1}_t)^\top) d{\bf B}^1_t - (\sigma^{Y,2}_t)^\top d{\bf B}^2_t \right).
\end{aligned}
\label{eq:Z:power:rewrite}
\end{equation}

To simplify the notations, let $\alpha=\{\alpha_t\}_{t\geq 0}\in\mathcal{P}_1\left(\mathbb{F}\right)$ and $\beta = \{\beta_t\}_{t\geq 0}\in \mathcal{P}_n\left(\mathbb{F}\right)$ be uniformly bounded processes, which are given by, for any $t\geq 0$,
\begin{equation*}
\beta_t=\Sigma_t^\top\hat{\pi}_t-\sigma^{Y,1}_t.
\end{equation*}
\begin{equation}
\label{eq:power:alpha:beta}
\alpha_t - (\lambda_t-\sigma^{Y,1}_t)^\top\beta_t =\lambda^\top_t\sigma^{Y,1}_t +\|\sigma^{Y,2}_t\|^2 -\mu^Y_t.
\end{equation}
Hence, the dynamics of $Z$ can be rewritten as, for any $t\geq 0$, 
\begin{equation}
\label{eq:Z:power}
dZ_t = p_t dt + Z_t\left(\alpha_t dt + \beta^\top_t d{\bf B}^1_t -(\sigma^{Y,2}_t)^\top d{\bf B}^2_t\right);
\end{equation}
the baseline investment strategy $\hat{\pi}$ can also be rewritten as, for any $t\geq 0$,
\begin{equation*}
\hat{\pi}_t=(\Sigma_t^\top)^{-1}(\sigma^{Y,1}_t+\beta_t).
\end{equation*}
Therefore, $Z$ satisfies \eqref{eq:Z:SPDE} with $\nu_t = p_t + Z_t\alpha_t$, $\kappa^1_t = Z_t\beta_t$, and $\kappa^2_t = -Z_t\sigma^{Y,2}_t$, for any $t\geq 0$. The coefficient $\beta$ measures the sensitivity of the exogenous fund value to salary ratio $Z$ with respect to the hedgeable risk ${\bf B}^1$ in the financial market.}\\

In the sequel, we shall use the notation $X^{\hat{\pi},0}=\{ X^{\hat{\pi},0}_t:=Z_t\}_{t\geq 0} $ to emphasize {\color{black}the} economic meaning {\color{black}of the minimum guarantee}, where the superscript $0$ indicates that the fund value starts at zero.

\subsection{Admissibility}\label{sec:power_admissibility}
With the stochastic domain given by, for any $t\geq 0$, $\mathcal{D}_t=\left(X^{\hat{\pi},0}_t,\infty\right)$, together with an integrability condition for constructing power forward utility preference, define the admissible set of investment strategies by
\begin{equation*}
\label{eq:A:power}
\mathcal{A} =  \{ \pi\in\mathcal{P}_n(\mathbb{F})  : X^{\pi}\pi \in \mathcal{L}^2_n,\  X^\pi_t>X^{\hat{\pi},0}_t, \   \text{for a.a. } (t,\omega) \in [0,\infty)\times \Omega   \}.
\end{equation*}
{\color{black} As being motivated above, 
t}he admissibility of an investment strategy $\pi$ ensures that the corresponding fund value to salary ratio must be strictly better than the baseline performance; that is, for any $t\geq 0$, $X^{\pi}_t>X_t^{\hat{\pi},0}$. Hence, the auxiliary process $X^{\hat{\pi},0}$ also renders a subsistence level which the worker's fund value to salary ratio $X$ must exceed; see, for example, \cite{method:financial:math}.\\

The following lemma first recalls a standard result for linear SDEs, which shall be used to show that $\hat{\pi}\in\mathcal{A}$ in the next proposition.
\begin{lemma} 
\label{lemma:sde}
Let $\{\tilde{W}_t\}_{t\geq 0} \in \mathcal{P}_d(\mathbb{F})$ be an $d$-dimensional Brownian motion. Let $\eta^1=\{\eta^1_t\}_{t\geq 0}\in \mathcal{P}_1(\mathbb{F})$, $\eta^2=\{\eta^2_t\}_{t\geq 0}\in \mathcal{P}_d(\mathbb{F})$, $\varphi^1=\{\varphi^1_t\}_{t\geq 0}\in \mathcal{P}_1(\mathbb{F})$, and $\varphi^2=\{\varphi^2_t\}_{t\geq 0}\in \mathcal{P}_d(\mathbb{F})$. Consider the following linear SDE: for any $t\geq 0$,
\begin{equation}
\label{eq:app:sde}
d\mathcal{X}_t = \eta^1_t dt + (\eta^2_t)^\top d\tilde{W}_t + \mathcal{X}_t\left(\varphi^1_t dt + (\varphi^2_t)^\top d\tilde{W}_t \right).
\end{equation}
If $\eta^1\in\mathcal{L}^2_1$, $\eta^2\in\mathcal{L}^2_d$, as well as $\varphi^1$ and $\varphi^2$ are both uniformly bounded, then there exists a unique solution $\mathcal{X}=\{\mathcal{X}_t\}_{t\geq 0}\in\mathcal{P}_1(\mathbb{F})$ of \eqref{eq:app:sde} such that $\mathcal{X}\in\mathcal{L}^2_1$. Moreover, $\mathcal{X}$ admits a continuous version.
\end{lemma}
\begin{proposition}\label{prop:non-empty}
We have $\hat{\pi}\in\mathcal{A}$, which is then non-empty.
\end{proposition}
\begin{proof}
By \eqref{eq:X} and \eqref{eq:power:alpha:beta}, $X^{\hat{\pi}}$ satisfies
\begin{equation*}
dX^{\hat{\pi}}_t = p_t dt +X^{\hat{\pi}}_t(\alpha_t dt + \beta^\top_t d{\bf B}^1_t -(\sigma^{Y,2}_t)^\top d{\bf B}^2_t).
\end{equation*}
Since $p_{\cdot}$, $\alpha$, $\beta$, and $\sigma^{Y,2}$ are all (uniformly) bounded, by Lemma \ref{lemma:sde}, $X^{\hat{\pi}}\in\mathcal{L}^2_1$, and hence $X^{\hat{\pi}}\hat{\pi}\in\mathcal{L}^2_n$ due to the uniform boundedness of $\hat{\pi}$. Since $X^{\hat{\pi}}$ and $X^{\hat{\pi},0}$ solve the same linear SDE (see \eqref{eq:Z:power}), while $X^{\hat{\pi}}_0=x_0>0=X^{\hat{\pi},0}_0$, by the comparison principle, $X^{\hat{\pi}}_t>X^{\hat{\pi},0}_t$ for a.a. $(t,\omega) \in [0,\infty)\times \Omega $. These show that $\hat{\pi}\in\mathcal{A}$.
\end{proof}

Define the set of investment strategies: \begin{align*}
\tilde{\mathcal{A}}=\Bigg\{\pi\in\mathcal{P}_n\left(\mathbb{F}\right):&\;\pi_t=\frac{X^{\hat{\pi},0}_t}{X^\pi_t} \hat{\pi}_t + \left(1-\frac{X^{\hat{\pi},0}_t}{X^\pi_t} \right) (\Sigma^\top_t)^{-1} (\sigma^{Y,1}_t+\xi_t),\;t\geq 0,\\&\;\text{for some uniformly bounded }\xi=\{\xi_t\}_{t\geq 0} \in \mathcal{P}_n(\mathbb{F})\Bigg\},
\end{align*}
in which each investment strategy is essentially a modification around the baseline strategy $\hat{\pi}$ with a uniformly bounded process $\xi$. The following proposition shows that the admissible set of investment strategies $\mathcal{A}$ is not only non-empty, but also rich enough.
\begin{proposition}
\label{pp:admissible}
We have $\tilde{\mathcal{A}}\subseteq\mathcal{A}$.
\end{proposition}

\begin{proof}
Let $\pi\in\mathcal{\tilde{A}}$, which can be simplified as, for any $t\geq 0$,
\begin{equation*}
\pi_t=(\Sigma^\top_t)^{-1}\left(\sigma^{Y,1}_t+\frac{X^{\hat{\pi},0}_t}{X^\pi_t} \beta_t + \left(1-\frac{X^{\hat{\pi},0}_t}{X^\pi_t} \right)\xi_t\right).
\end{equation*}
Also, let $\Tilde{X}_t := X^\pi_t-X^{\hat{\pi},0}_t$, for $t\geq 0$. We have, for any $t\geq 0$,
\begin{equation}
\begin{aligned}
d\Tilde{X}_t = & \ \Tilde{X}_t \Big(  \left(  \xi^\top_t(\lambda_t-\sigma^{Y,1}_t) + \lambda^\top_t\sigma^{Y,1}_t+\|\sigma^{Y,2}_t\|^2- \mu^Y_t  \right)  dt  \\ & \ \quad\quad+ \xi_t^\top d{\bf B}^1_t  -  (\sigma^{Y,2}_t)^\top d{\bf B}^2_t\Big).
\end{aligned}
\label{eq:tilde:X}
\end{equation}
By the (uniform) boundedness and Lemma \ref{lemma:sde}, $\Tilde{X}$ satisfying \eqref{eq:tilde:X} implies that $\Tilde{X}\in\mathcal{L}^2_1$. As $X^{\hat{\pi},0}\in\mathcal{L}^2_1$ (see the proof of Proposition \ref{prop:non-empty}), $X^\pi\equiv\Tilde{X}+X^{\hat{\pi},0}\in\mathcal{L}^2_1$.\\

Moreover, since $\tilde{X}_0 = x_0>0$, for any $t\geq 0$,
\begin{equation*}
\tilde{X}_t = x_0e^{\int_0^t \left(  \xi^\top_s(\lambda_s-\sigma^{Y,1}_s) + \lambda^\top_s\sigma^{Y,1}_s+\|\sigma^{Y,2}_s\|^2- \mu^Y_s  \right)ds   } \mathcal{E}_t,
\end{equation*}
where the Dol{\'e}ans-Dade exponential $\mathcal{E}=\{\mathcal{E}_t\}_{t\geq 0}$ satisfies that $\mathcal{E}_0=1$ and, for any $t\geq 0$,
\begin{equation*}
d\mathcal{E}_t = \mathcal{E}_t\left( \xi_t^\top d{\bf B}^1_t - (\sigma_t^{Y,2})^\top d{\bf B}^2_t \right).
\end{equation*}
Obviously, $\tilde{X}_t>0$, and thus $X^\pi_t>X^{\hat{\pi},0}_t$, $\text{for a.a. } (t,\omega) \in [0,\infty)\times \Omega$.\\

Finally, since $p_{\cdot}\in\mathbb{R}_+$ and $X^{\hat{\pi},0}$ solves the linear SDE \eqref{eq:Z:power}, by the comparison principle, $X^{\hat{\pi},0}_t>0$ for $t\geq 0$. Again, by the (uniform) boundedness, together with $X^{\hat{\pi},0}_t/X^\pi_t\in[0,1)$ for $t\geq 0$, $\pi$ is also uniformly bounded. Therefore, $X^\pi\pi\in\mathcal{L}^2_n$. 



\end{proof}


\subsection{Non-Zero Volatility Forward Preferences}\label{sec:admis_power} 
Inspired by {\color{black}Sections \ref{sec:SPDE} and  \ref{sec:power_admissibility}}, the worker's forward utility preference is driven by the comparison on the performance between her investment strategy and the baseline strategy. Moreover, the preference depends on the worker's appetite with respect to the hedgeable and non-hedgeable risks. The following theorem constructs such a (non-)zero volatility power forward utility preference of the worker, together with her corresponding optimal investment strategy{\color{black}; its proof is relegated to Appendix \ref{sec:app:proof:power}.}
\begin{theorem}\label{pp:power}
Let $V=\{V_t\}_{ t\geq 0} \in \mathcal{P}_1(\mathbb{F})$ be a process, given by $V_0=0$ and, for any $t\geq 0$, 
    \begin{equation*}
    \label{eq:V:power:1}
        dV_t = v_t dt + (\theta_t^1)^\top d{\bf B}^1_t + (\theta_t^2)^\top d{\bf B}^2_t,  
    \end{equation*}
where $v = \{v_t\}_{t\geq 0}\in \mathcal{P}_1(\mathbb{F})$, $\theta^1 = \{\theta^1_t\}_{t\geq 0}\in \mathcal{P}_n(\mathbb{F})$, and $\theta^2 =\{\theta^2_t\}_{t\geq 0} \in \mathcal{P}_m(\mathbb{F})$ are uniformly bounded processes, such that, for any $t\geq 0$, 
\begin{equation}
\begin{aligned}
v_t = &  -\frac{\gamma(1+\gamma)}{2}( \|\sigma^{Y,1}_t\|^2 +  \|\sigma^{Y,2}_t\|^2)  - \frac{\gamma\| \lambda_t - \gamma \sigma^{Y,1}_t + \theta^1_t  \|^2}{2(1-\gamma)} \\
& + \gamma\left( (\theta_t^1)^\top \sigma^{Y,1}_t +  (\theta_t^2)^\top \sigma^{Y,2}_t +\mu^Y_t \right) - \frac{\|\theta_t^1\|^2+\|\theta_t^2\|^2}{2}.
\end{aligned}
\label{eq:V:power:2}
\end{equation}

The random field, for any $t\geq 0$ and $x\in\mathbb{R}$,
\begin{equation}
\label{eq:U:power}
U(x,t) = \begin{cases}
\frac{1}{\gamma} (x-X^{\hat{\pi},0}_t)^\gamma  e^{V_t} &\text{if } x> X^{\hat{\pi},0}_t\\
-\infty &\text{otherwise}
\end{cases},
\end{equation}
is a power forward utility preference on the fund value to salary ratio. In addition, the volatility processes of its translated random field $\{\tilde{U}(\tilde{x},t)=U(\tilde{x}+X^{\hat{\pi},0}_t,t)\}_{ \tilde{x}\in \mathbb{R}_{++} ,  t\geq 0 }$ are given by, for any $\Tilde{x}\in \mathbb{R}_{++}$ and $t\geq 0$,
\begin{equation}
\label{eq:vol:power}
a_1(\tilde{x},t) = \frac{\Tilde{x}^\gamma}{\gamma}e^{V_t}\theta^1_t\quad \text{and} \quad a_2(\tilde{x},t) = \frac{\Tilde{x}^\gamma}{\gamma}e^{V_t}\theta^2_t.
\end{equation}
Moreover, the optimal investment strategy is given by, for any $t\geq 0$, 
\begin{equation}
\label{eq:pi*:power:1}
\pi_t^* = \frac{X^{\hat{\pi},0}_t}{X^*_t} \hat{\pi}_t + \left( 1-\frac{X^{\hat{\pi},0}_t}{X^*_t}\right) (\Sigma^\top_t)^{-1}\frac{\lambda_t-\gamma\sigma^{Y,1}_t + \theta^1_t}{1-\gamma},
\end{equation}
where $X^*:\equiv X^{\pi^*}$  satisfies, for any $t\geq 0$, 
\begin{align*}
    dX^*_t =& 
 \  \Bigg(  p_t + X^{\hat{\pi},0}_t\left( (\lambda_t-\sigma^{Y,1}_t)^\top\beta_t + \lambda^\top_t \sigma^{Y,1}_t + \|\sigma^{Y,2}_t\|^2 -\mu^Y_t \right)   \\
 & \quad + (X_t^*-X^{\hat{\pi},0}_t) \left( \frac{(\lambda_t-\sigma^{Y,1}_t)^\top(\lambda_t-\sigma^{Y,1}_t+\theta^1_t)}{1-\gamma} +  \lambda^\top_t \sigma^{Y,1}_t + \|\sigma^{Y,2}_t\|^2 -\mu^Y_t  \right)\Bigg)dt  
 \\ & \ + \Bigg( X^{\hat{\pi},0}_t \beta_t  +   (X^*_t-X^{\hat{\pi},0}_t)\left( \frac{\lambda_t- \sigma^{Y,1}_t + \theta^1_t }{1-\gamma}  \right)         \Bigg)^\top       d{\bf B}_t^1 
- X_t^*(\sigma^{Y,2}_t)^\top d{\bf B}^2_t.
\end{align*} 
\end{theorem}


{\color{black} When managing her pension fund, t}he worker observes by how much more her fund value to salary ratio exceeding the chosen baseline ratio, or equivalently, by how much more her absolute fund value exceeding the baseline absolute fund value, using her salary as the num{\'e}raire. The worker's forward preference is then given by this surplus being evaluated by the classical power utility function, together with multiplying a homothetic factor. The worker also exogenously chooses the {\color{black}components of the} volatility processes $\theta^1$ and $\theta^2$, with respect to the hedgeable risk ${\bf B}^1$ in the financial market and the non-hedgeable risk ${\bf B}^2$ in the labour market, governing the evolution of the homothetic factor to her preference.  In particular, if the volatility processes are both exogenously chosen to be zero, the worker's translated forward preference is also of zero volatility; in this case, the time monotonicity of her translated forward preference depends on the sign of the deterministic $v_{\cdot}$ (with $\theta^1\equiv\theta^2\equiv 0$). {\color{black}In general, the volatility processes inherent the structure of the preferences themselves while allow the worker to exogenously choose $\theta^1$ and $\theta^2$ for the homothetic factor which pass on to the volatility processes $a_1$ and $a_2$ of her preferences. In this case, the larger the worker's preference, or the volatility $\theta^1$ or $\theta^2$ of the homothetic factor, is, the more volatile the worker's preference is.}\\

The worker's optimal investment strategy, based on her power forward utility preference, as outlined in Theorem \ref{pp:power}, is clearly a convex combination of the baseline investment strategy $\hat{\pi}$ and a myopic strategy $(\Sigma^\top_t)^{-1}\frac{\lambda_t-\gamma\sigma^{Y,1}_t + \theta^1_t}{1-\gamma}$, for $t\geq 0$. This myopic strategy is in fact consistent with the optimal investment strategy in the classical backward setting of \cite{CAIRNS2006843:stochastic:lifestyle}, in the case of $p_{\cdot}\equiv 0$ therein (see Cases 1 and 2), as well as $\theta^1\equiv 0$ endogenously from the backward value function. {\color{black} Thus, the factor $\theta^1$ can also be viewed as a distortion parameter for the backward myopic strategy.} Unlike the backward approach in \cite{CAIRNS2006843:stochastic:lifestyle}, where the worker's optimal investment strategy at a particular time depends on the projection of her salary contribution  afterwards (see Cases 3 and 4 therein, {\color{black} where the former has been discussed in Section \ref{sec:backward:pitfall}}), her optimal investment strategy herein at that time, based on the forward approach, depends {\color{black} only} on her past salary contribution {\color{black} via the baseline fund value to salary ratio. Hence, her optimal strategy at a certain time under the forward approach is independent of the {\color{black} model of the} environment in the future. This is substantially different from the backward approach, where the worker's optimal strategy depends on the model of the future environment via the modelled projections of the future salary in that case. Thus, the forward approach is capable of any potential model updates than its backward counterpart, see Section \ref{sec:mot_eg_revisited} for a detailed discussion.} 
Moreover, observe that the convex combination coefficient is the quotient of the two ratios for the fund values (of baseline and optimality) to salary, which is actually the ratio of the baseline and optimal absolute fund values. Therefore, the worker's optimal investment strategy does not directly depend on the salary value, but it depends implicitly from the absolute fund value dynamics and the coefficient $\sigma^{Y,1}$ of the salary value's dynamics. However, the worker's forward preference indeed evaluates her fund value to salary ratio, in order for the (super-)martingale conditions to be satisfied. Since the ratio of the baseline and optimal absolute fund values is the convex combination coefficient, her optimal investment strategy tends to last in either one of the baseline and optimal investment strategies, unless the other one outperforms the lasting one substantially at a future time. For example, when the baseline and optimal absolute fund values are close, the worker's optimal investment strategy leans towards the baseline strategy, and vice versa; see also the numerical example in the next section.

\subsection{Numerical Example}
\label{sec:power:numeric}
This section further illustrates the worker's power forward utility preference and optimal investment strategy in Theorem \ref{pp:power} via a numerical example. {\color{black}Consider the same set of constant parameters as in Section \ref{sec:backward:pitfall}, with the exceptions that $\sigma^{Y,2}_{\cdot}=5\%$ and $m=1$. Again, the volatility $\sigma^{Y,2}_{\cdot}$ of her salary with respect to the non-hedgeable risk is set to be less than $10\%$ to reflect that her salary is less volatile than the risky asset. Since $n=m=1$, denote the independent one-dimensional Brownian motions ${\bf B}^1$ and ${\bf B}^2$ simply as $B^1$ and $B^2$. Utilizing her power forward utility preference, the worker chooses $\theta^1_{\cdot}=0$ and $\theta^2_{\cdot}=0.2$ as the constant parameters for the volatility processes of her translated forward utility preferences; that is, her forward preference is insensitive to the hedgeable risk but it is moderately responsive to the non-hedgable one.}\\

\begin{figure}[!h]
\centering
\begin{subfigure}{.33\textwidth}
\centering
\includegraphics[scale=0.3]{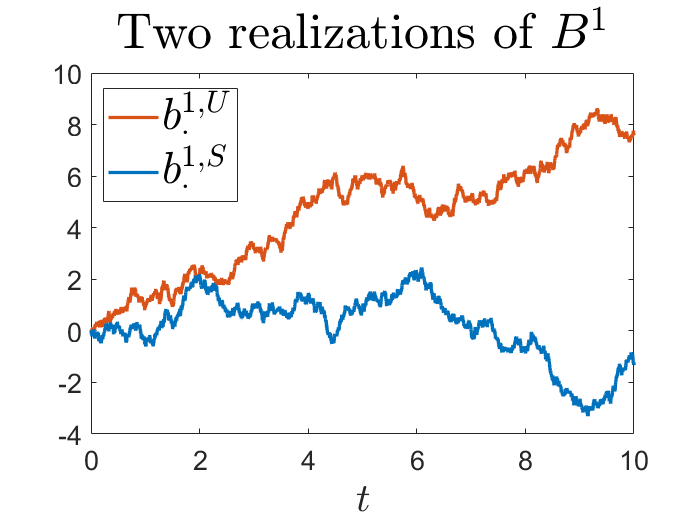}
\caption{Two realizations of $B^1$}
\label{fig:B1}
\end{subfigure}%
\begin{subfigure}{.33\textwidth}
\centering
\includegraphics[scale=0.3]{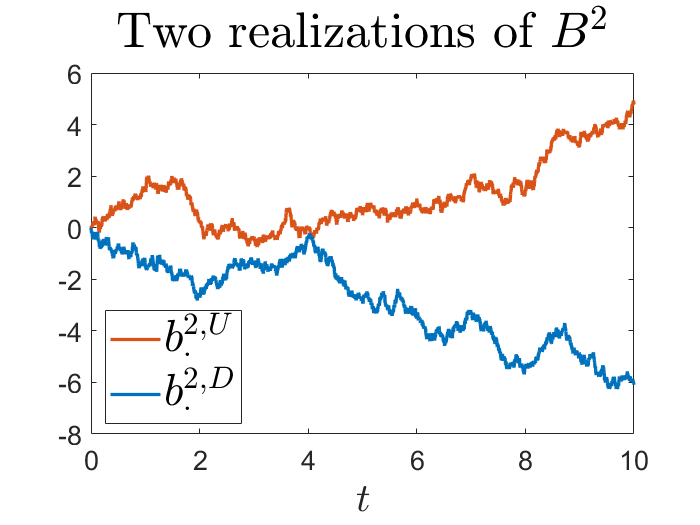}
\caption{Two realizations of $B^2$}
\label{fig:B2}
\end{subfigure}%
\begin{subfigure}{.34\textwidth}
\centering
\includegraphics[scale=0.3]{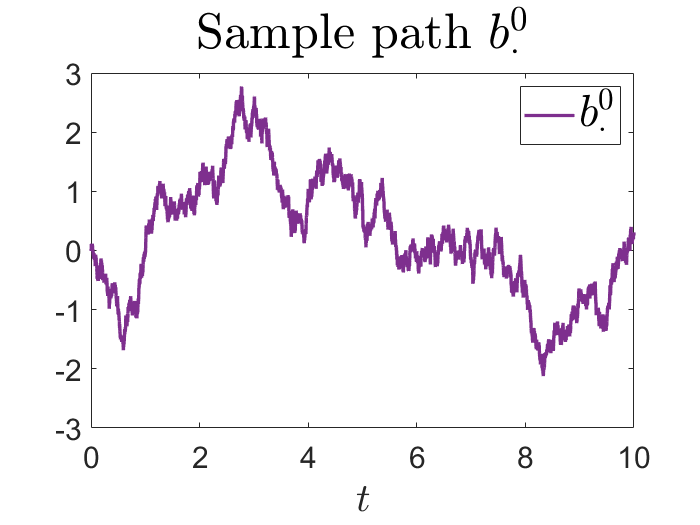}
\caption{Sample path $b^0_{\cdot}$}
\label{fig:B3}
\end{subfigure}%
\caption{Realizations of Brownian motions}
\label{fig:B}
\end{figure}

This example particularly sheds light on the stochasticity of her forward utility preference and optimal investment strategy {\color{black}solved at the current time $t=0$}. To this end, consider four scenarios $\omega_1,\omega_2,\omega_3,\omega_4$, which yield the following sample paths of the Brownian motions: for any $t\geq 0$,
\begin{align*}
\left(B^1_t,B^2_t\right)\left(\omega_1\right)=&\;\left(b^{1,U}_t,b^0_t\right),\\
\left(B^1_t,B^2_t\right)\left(\omega_2\right)=&\;\left(b^{1,S}_t,b^0_t\right),\\
\left(B^1_t,B^2_t\right)\left(\omega_3\right)=&\;\left(b^0_t,b^{2,U}_t\right),\\
\left(B^1_t,B^2_t\right)\left(\omega_4\right)=&\;\left(b^0_t,b^{2,D}_t\right),
\end{align*}
where $b^{1,U}_{\cdot},b^{1,S}_{\cdot},b^{2,U}_{\cdot},b^{2,D}_{\cdot},b^{0}_{\cdot}$ are depicted in Figure \ref{fig:B} (in which $t=10$ is a time before the accumulation period ends). Under scenarios $\omega_1$ and $\omega_2$, the non-hedgeable risk $B^2$ evolves moderately around zero as in Figure \ref{fig:B3}, and the hedgeable risk $B^1$ moves up in the first scenario while it remains relatively stable in the second scenario (see Figure \ref{fig:B1}). Similar interpretations under the scenarios $\omega_3$ and $\omega_4$, with $B^1$ and $B^2$ as in Figures \ref{fig:B3} and \ref{fig:B2} respectively.\\

This example also considers two baseline fund value to salary ratios, with $\beta\equiv {\color{black}0.25}$ and $\beta\equiv {\color{black}-0.25}$ respectively for the sensitivity of the subsistence level with respect to the hedgeable risk $B^1$, and studies their influences on the worker's forward utility preference and optimal investment strategy.

\subsubsection{Forward Utility Preferences}
\begin{figure}[!h]
\centering
\begin{subfigure}{.5\textwidth}
\centering
\includegraphics[scale=0.4]{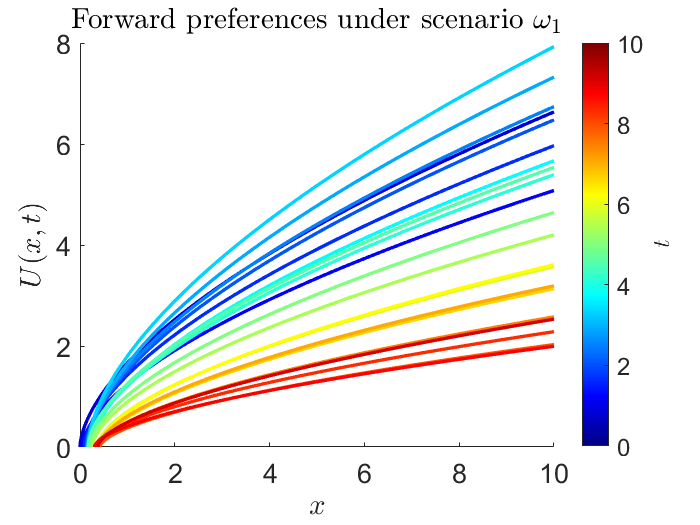}
\caption{$\omega=\omega_1$, $\beta\equiv {\color{black}-0.25}$}
\label{fig:power:U:UM:B-05}
\end{subfigure}%
\begin{subfigure}{.5\textwidth}
\centering
\includegraphics[scale=0.4]{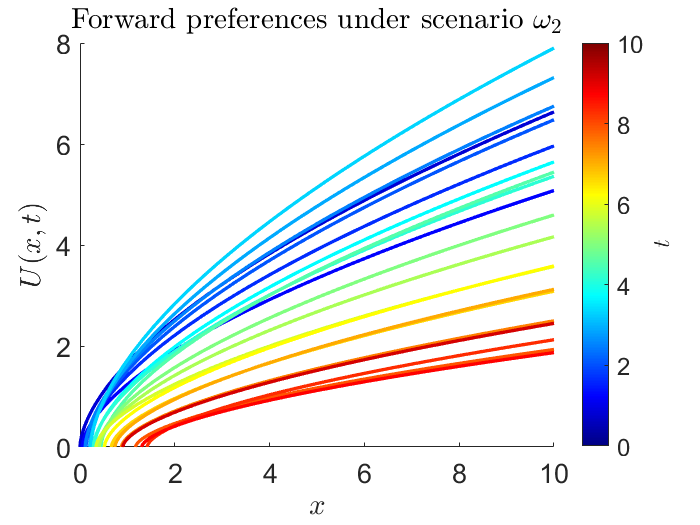}
\caption{$\omega=\omega_2$, $\beta\equiv {\color{black}-0.25}$}
\label{fig:power:U:DM:B-05}
\end{subfigure}%

\begin{subfigure}{.5\textwidth}
\centering
\includegraphics[scale=0.4]{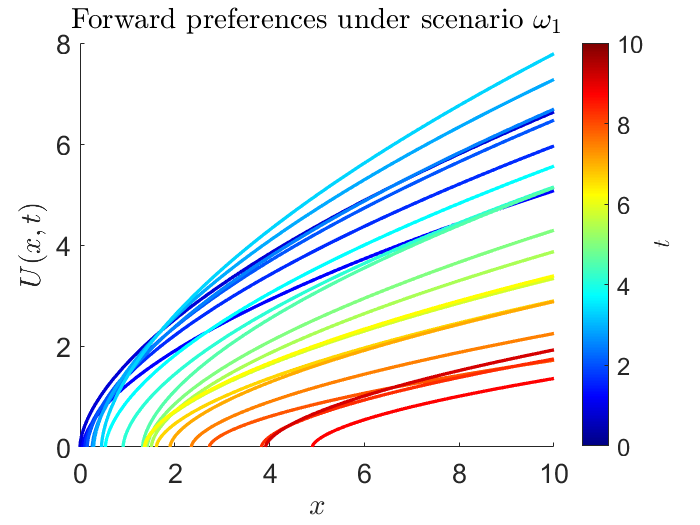}
\caption{$\omega=\omega_1$, $\beta\equiv {\color{black}0.25}$}
\label{fig:power:U:UM:B05}
\end{subfigure}%
\begin{subfigure}{.5\textwidth}
\centering
\includegraphics[scale=0.4]{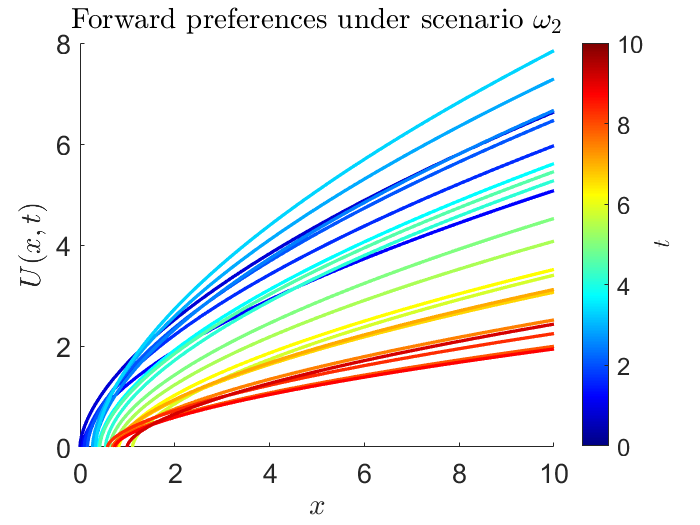}
\caption{$\omega=\omega_2$, $\beta\equiv {\color{black}0.25}$}
\label{fig:power:U:DM:B05}
\end{subfigure}%

\caption{Realizations of worker's forward utility preferences under scenarios $\omega_1$ and $\omega_2$ and with respect to two different baseline performances}
\label{fig:power:utility:market}
\end{figure}

\begin{figure}[!h]
\centering
\begin{subfigure}{.5\textwidth}
\centering
\includegraphics[scale=0.4]{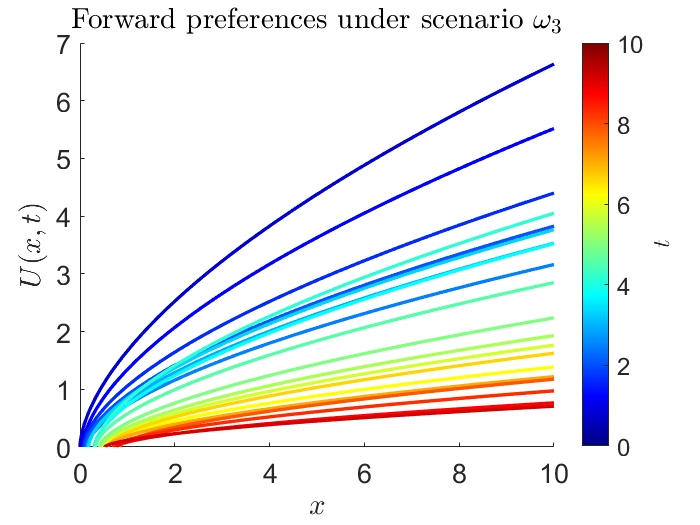}
\caption{$\omega=\omega_3$, $\beta\equiv {\color{black}-0.25}$}
\label{fig:power:U:US:B-05}
\end{subfigure}%
\begin{subfigure}{.5\textwidth}
\centering
\includegraphics[scale=0.4]{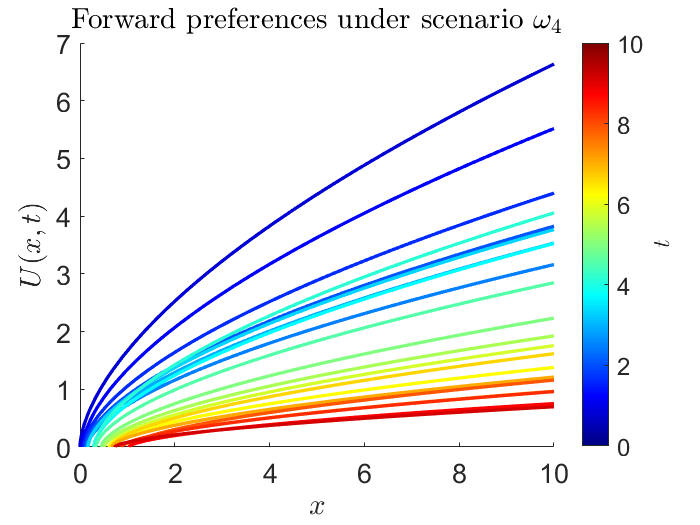}
\caption{$\omega=\omega_4$, $\beta\equiv {\color{black}-0.25}$}
\label{fig:power:U:SS:B-05}
\end{subfigure}%

\begin{subfigure}{.5\textwidth}
\centering
\includegraphics[scale=0.4]{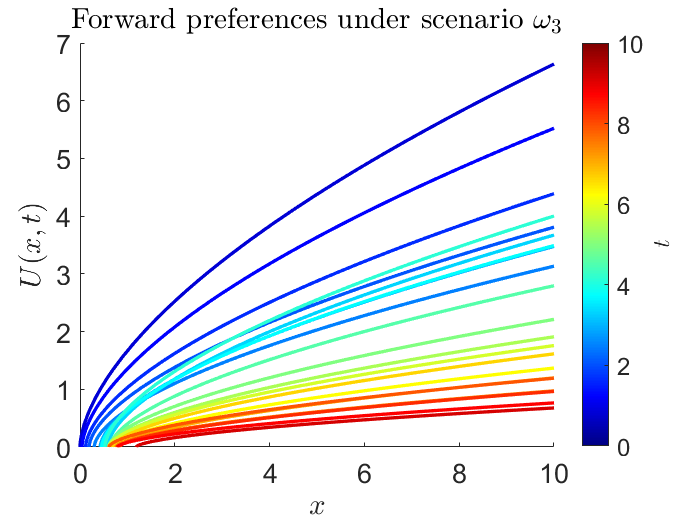}
\caption{$\omega=\omega_3$, $\beta\equiv {\color{black}0.25}$}
\label{fig:power:U:US:B05}
\end{subfigure}%
\begin{subfigure}{.5\textwidth}
\centering
\includegraphics[scale=0.4]{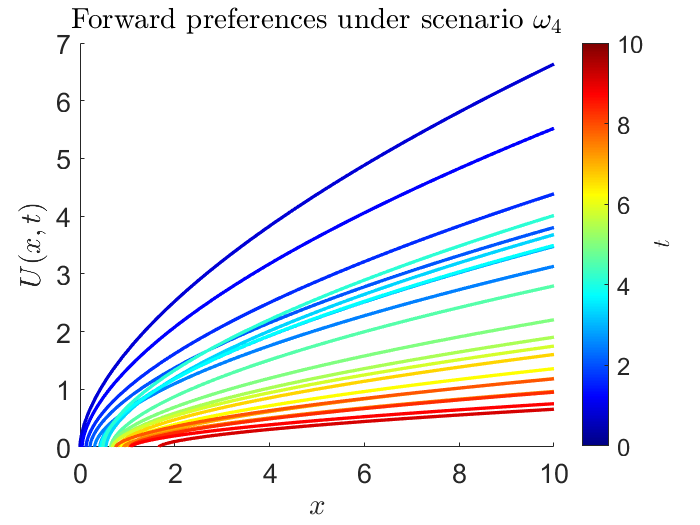}
\caption{$\omega=\omega_4$, $\beta\equiv {\color{black}0.25}$}
\label{fig:power:U:SS:B05}
\end{subfigure}%

\caption{Realizations of worker's forward utility preferences under scenarios $\omega_3$ and $\omega_4$ and with respect to two different baseline performances}
\label{fig:power:utility:salary}
\end{figure}

The worker's forward utility preferences are summarized in Figures \ref{fig:power:utility:market} and \ref{fig:power:utility:salary}. Her preferences $U\left(\cdot,t;\omega\right)$ are pictured as functions of the ratio argument $x\in\mathbb{R}$, for any fixed scenario $\omega\in\Omega$ and time $t\geq 0$. Their time dependence are illustrated by various colors highlighting the respective functions. Figure \ref{fig:power:utility:market} depicts the scenarios $\omega_1$ and $\omega_2$, while Figure \ref{fig:power:utility:salary} illustrates the scenarios $\omega_3$ and $\omega_4$.\\

\begin{figure}[!h]
    \centering
    \begin{subfigure}{.5\textwidth}
    \centering
    \includegraphics[scale=0.4]{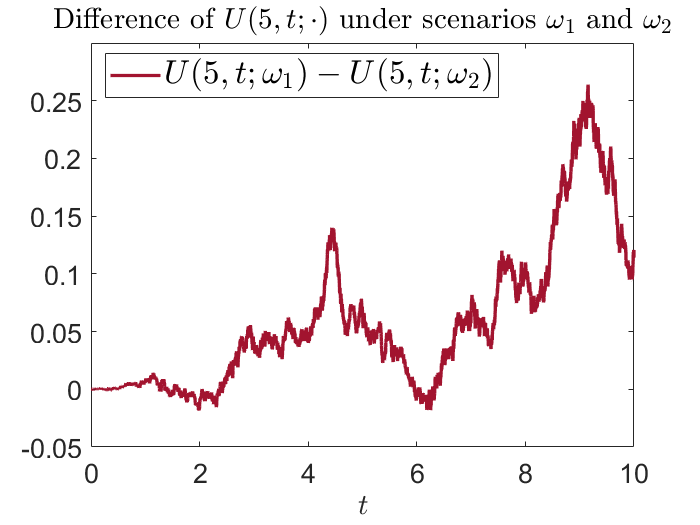}
    \caption{$\beta\equiv {\color{black}-0.25}$}
    \label{fig:power:DU:M:Bm05}
    \end{subfigure}%
    \begin{subfigure}{.5\textwidth}
    \centering
    \includegraphics[scale=0.4]{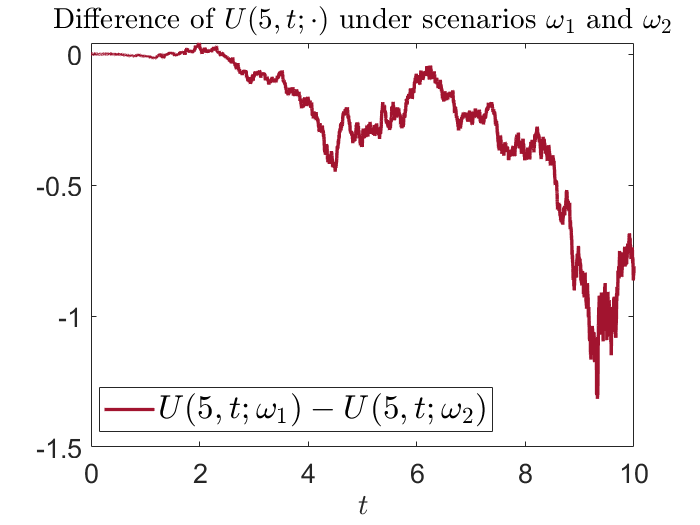}
    \caption{$\beta\equiv {\color{black}0.25}$}
    \label{fig:power:DU:M:B05}
    \end{subfigure}%
    \caption{Difference of realized worker's forward utility preferences, between scenarios $\omega_1$ and $\omega_2$, with respect to two different baseline performances, and at ratio argument $x=5$}
    \label{fig:power:du:M}
\end{figure}

If the worker compares the baseline performance with $\beta\equiv {\color{black}-0.25}$, contrasting between Figures \ref{fig:power:U:UM:B-05} and \ref{fig:power:U:DM:B-05}, her preference under the scenario $\omega_1$ is in general larger than that under the scenario $\omega_2$; see also Figure \ref{fig:power:DU:M:Bm05} for the preference value difference between the two scenarios fixing the ratio argument $x=5$. With the sensitivity of the baseline ratio with respect to the hedgeable risk $B^1$ being negative, when $B^1$ moves up in the first scenario, it draws down the baseline ratio. Therefore, the x-intercepts, which are the baseline ratio values, in Figure \ref{fig:power:U:UM:B-05} are smaller than those in Figure \ref{fig:power:U:DM:B-05}, and thus leading to her generally larger preference values in the first scenario. If the worker compares the baseline performance with $\beta\equiv {\color{black}0.25}$, the opposite holds between Figures \ref{fig:power:U:UM:B05} and \ref{fig:power:U:DM:B05} (see also Figure \ref{fig:power:DU:M:B05} for their difference), since the sensitivity of the baseline ratio with respect to the hedgeable risk $B^1$ is positive.\\

\begin{figure}[!h]
    \centering
    \begin{subfigure}{.5\textwidth}
    \centering
    \includegraphics[scale=0.4]{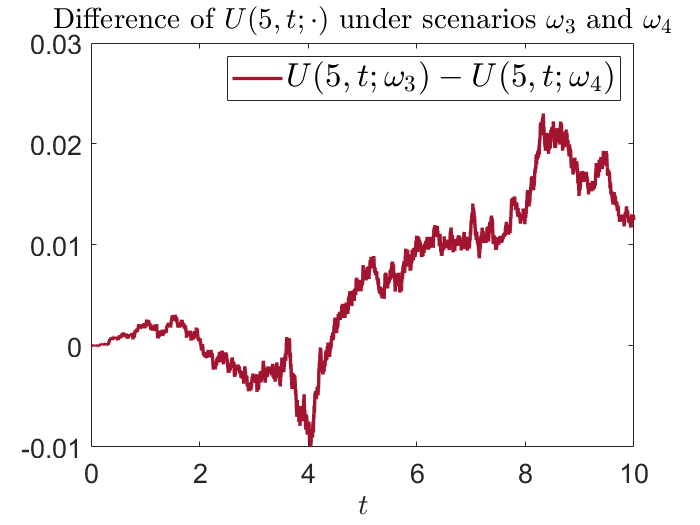}
    \caption{$\beta\equiv {\color{black}-0.25}$}
    \label{fig:power:DU:Bm05}
    \end{subfigure}%
    \begin{subfigure}{.5\textwidth}
    \centering
    \includegraphics[scale=0.4]{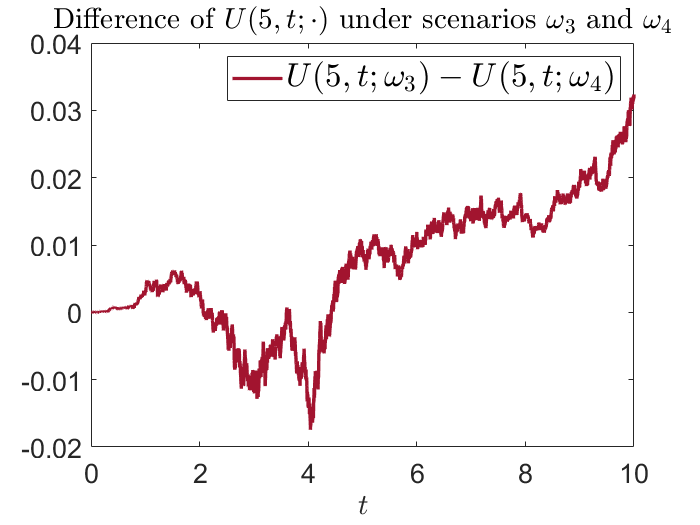}
    \caption{$\beta\equiv {\color{black}0.25}$}
    \label{fig:power:DU:B05}
    \end{subfigure}%
    \caption{Difference of realized worker's forward utility preferences, between scenarios $\omega_3$ and $\omega_4$, with respect to two different baseline performances, and at ratio argument $x=5$}
    \label{fig:power:du}
\end{figure}

From Figure \ref{fig:power:utility:salary}, regardless of which baseline performances the worker compares to, as the sensitivity of the baseline ratios with respect to the non-hedgeable risk $B^2$ is negative (which is given by {\color{black}$-\sigma^{Y,2}_{\cdot}=-0.05$}), the baseline ratio values as the x-intercepts in the scenario $\omega_4$ (when the non-hedgeable risk $B^2$ moves down, c.f. Figures \ref{fig:power:U:SS:B-05} and \ref{fig:power:U:SS:B05}) are larger than those in the scenario $\omega_3$ (when the non-hedgeable risk $B^2$ moves up, c.f. Figures \ref{fig:power:U:US:B-05} and \ref{fig:power:U:US:B05}). Hence, the worker's preference values are in general larger in the third scenario; see also Figure \ref{fig:power:du} for the preference value differences when $x=5$.

\subsubsection{Optimal Investment Strategy}
\begin{figure}[!h]
\centering
\begin{subfigure}{.5\textwidth}
\centering
\includegraphics[scale=0.4]{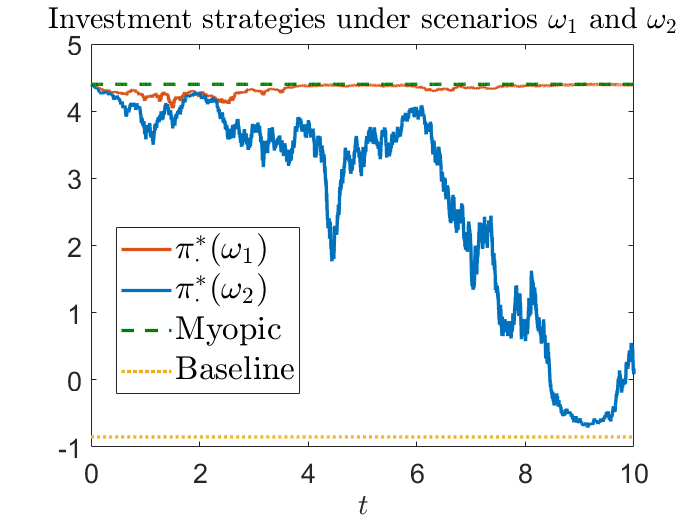}
\caption{$\beta\equiv {\color{black}-0.25}$}
\label{fig:power:Pi:M:b-05}
\end{subfigure}%
\begin{subfigure}{.5\textwidth}
\centering
\includegraphics[scale=0.4]{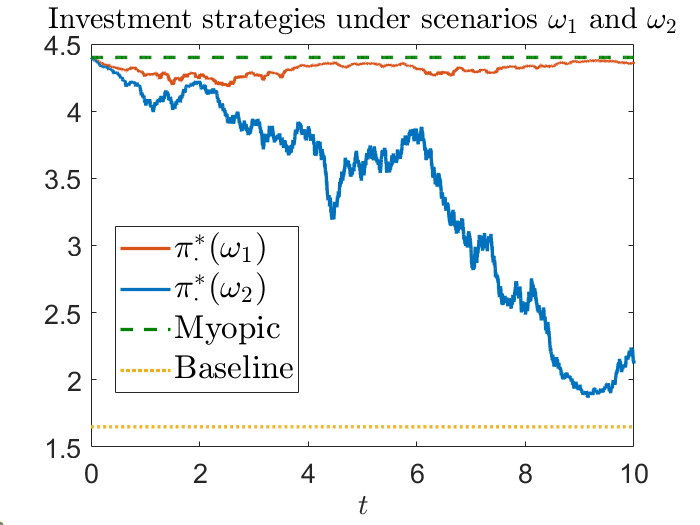}
\caption{$\beta\equiv {\color{black}0.25}$}
\label{fig:power:Pi:M:b05}
\end{subfigure}%

\caption{Realizations of worker's optimal investment strategies under scenarios $\omega_1$ and $\omega_2$ and with respect to two different baseline performances}
\label{fig:power:Pi:M}
\end{figure}

\begin{figure}[!h]
\centering
\begin{subfigure}{.5\textwidth}
\centering
\includegraphics[scale=0.4]{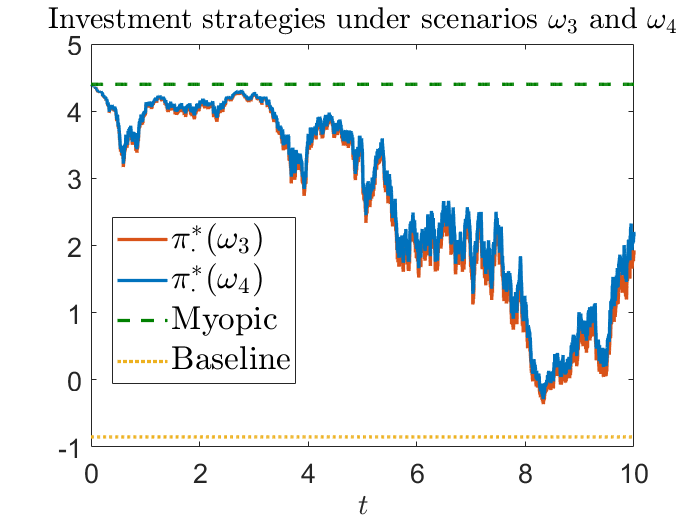}
\caption{$\beta\equiv {\color{black}-0.25}$}
\label{fig:power:Pi:S:b-05}
\end{subfigure}%
\begin{subfigure}{.5\textwidth}
\centering
\includegraphics[scale=0.4]{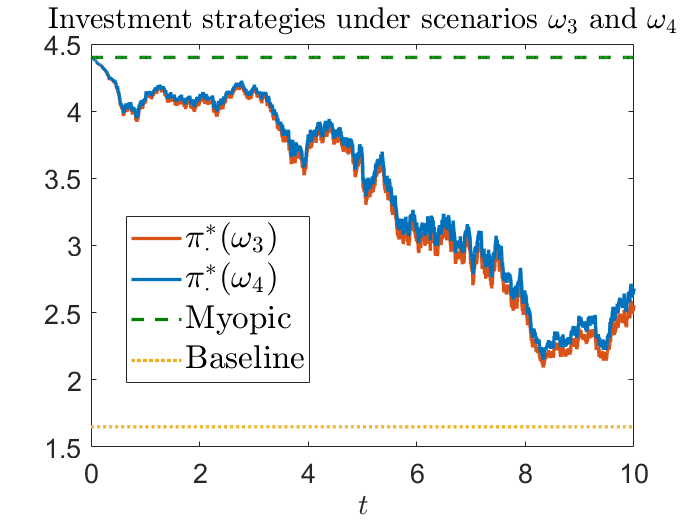}
\caption{$\beta\equiv {\color{black}0.25}$}
\label{fig:power:Pi:S:b05}
\end{subfigure}%

\caption{Realizations of worker's optimal investment strategies under scenarios $\omega_3$ and $\omega_4$ and with respect to two different baseline performances}
\label{fig:power:Pi:S}
\end{figure}

The worker's optimal investment strategies are summarized in Figures \ref{fig:power:Pi:M} and \ref{fig:power:Pi:S}; Figure \ref{fig:power:Pi:M} illustrates the scenarios $\omega_1$ and $\omega_2$, while Figure \ref{fig:power:Pi:S} depicts the scenarios $\omega_3$ and $\omega_4$. Under the given parameters, the myopic strategy $(\Sigma^\top_{\cdot})^{-1}\frac{\lambda_{\cdot}-\gamma\sigma^{Y,1}_{\cdot} + \theta^1_{\cdot}}{1-\gamma}={\color{black}4.4}$; the exogenous baseline strategy $\hat{\pi}_{\cdot}\equiv {\color{black}-0.85}<0$ when $\beta\equiv {\color{black}-0.25}$, while $\hat{\pi}_{\cdot}\equiv {\color{black}1.65}>0$ when $\beta\equiv {\color{black}0.25}$. Recall that her optimal investment strategy is a convex combination of the baseline and myopic strategies, with the coefficient for the baseline strategy given by the quotient of two ratios for the fund values, respectively of baseline and of optimality, to salary. Since the initial baseline fund value is zero, her optimal strategy starts from the myopic one.\\

\begin{figure}[!h]
\centering
\begin{subfigure}{.5\textwidth}
\centering
\includegraphics[scale=0.4]{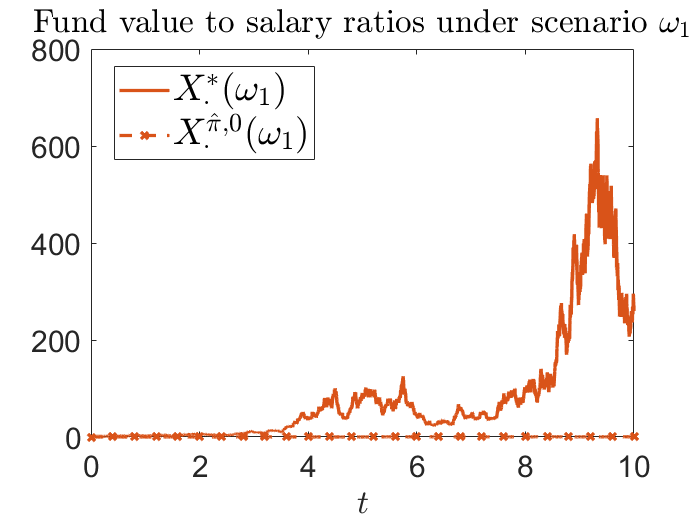}
\caption{$\beta\equiv {\color{black}-0.25}$}
\label{fig:power:XZ:UM:b-05}
\end{subfigure}%
\begin{subfigure}{.5\textwidth}
\centering
\includegraphics[scale=0.4]{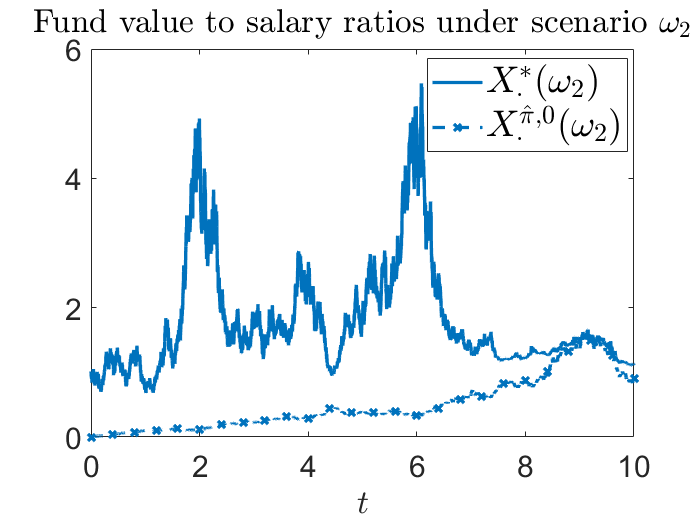}
\caption{$\beta\equiv {\color{black}-0.25}$}
\label{fig:power:XZ:DM:b-05} 
\end{subfigure}%

\begin{subfigure}{.5\textwidth}
\centering
\includegraphics[scale=0.4]{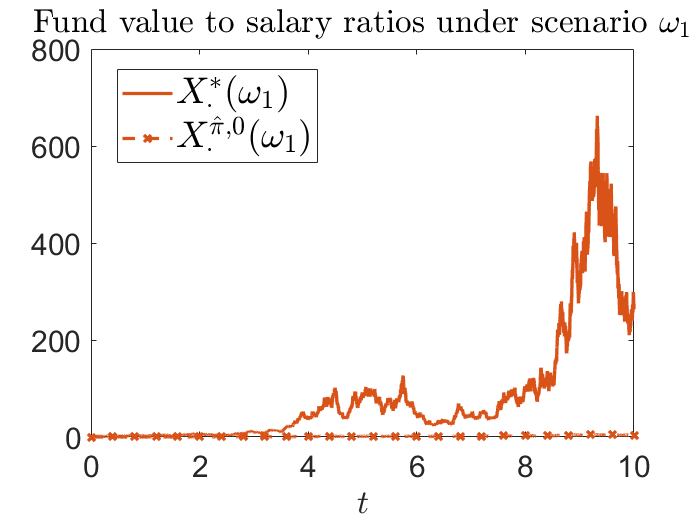}
\caption{$\beta\equiv {\color{black}0.25}$}
\label{fig:power:XZ:UM:b05}
\end{subfigure}%
\begin{subfigure}{.5\textwidth}
\centering
\includegraphics[scale=0.4]{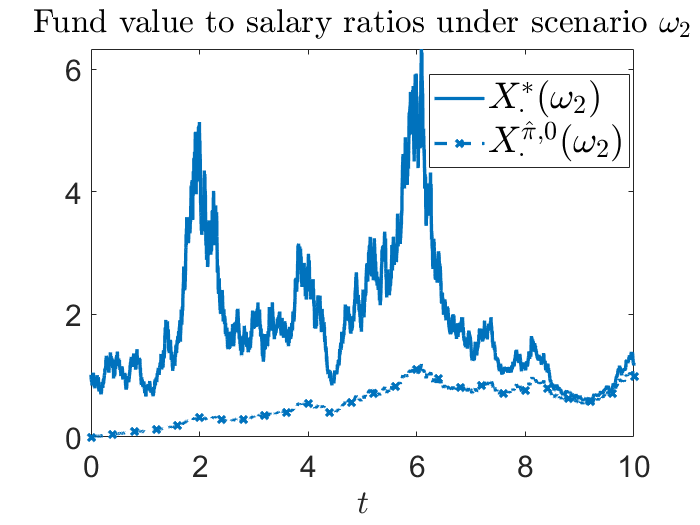}
\caption{$\beta\equiv {\color{black}0.25}$}
\label{fig:power:XZ:DM:b05} 
\end{subfigure}%

\caption{Realizations of worker's pension fund value to salary ratios under scenarios $\omega_1$ and $\omega_2$ (be mindful for the scale difference) and with respect to two different baseline performances}
\label{fig:power:XZ:M}
\end{figure}

From Figure \ref{fig:power:Pi:M}, the worker's optimal investment strategy follows the myopic strategy closely when the hedgeable risk $B^1$ moves up in the first scenario, while she tends to follow the baseline investment strategy as time propagates when $B^1$ remains relatively stable in the second scenario, no matter which baseline performances she compares to. Under the scenario $\omega_1$, the upward movement of $B^1$ drives up the risky asset value as well as the salary. On one hand, by following the positive myopic strategy, the worker's pension fund value grows substantially more than her own salary, leading to a large ratio value; on the other hand, by the exogenous baseline strategy, the performance in terms of the fund value to salary ratio stays relatively moderate. Therefore, the difference between the two ratios widens as time progresses (see Figures \ref{fig:power:XZ:UM:b-05} and \ref{fig:power:XZ:UM:b05}), and thus the coefficient for the baseline strategy is negligible; this explains why the worker follows the myopic strategy in this case. Under the scenario $\omega_2$, relatively stable hedgeable risk $B^1$ does not enlarge the difference between the two ratios much, but instead, the ratios gradually evolve adjacent to each other moving forward in time (see Figures \ref{fig:power:XZ:DM:b-05} and \ref{fig:power:XZ:DM:b05}), and thus the coefficient for the myopic strategy eventually becomes negligible; these illustrate why the worker's optimal investment strategies diverge from the myopic one while converge to the baseline strategies in this case.

\begin{figure}[!h]
\centering
\begin{subfigure}{.5\textwidth}
\centering
\includegraphics[scale=0.4]{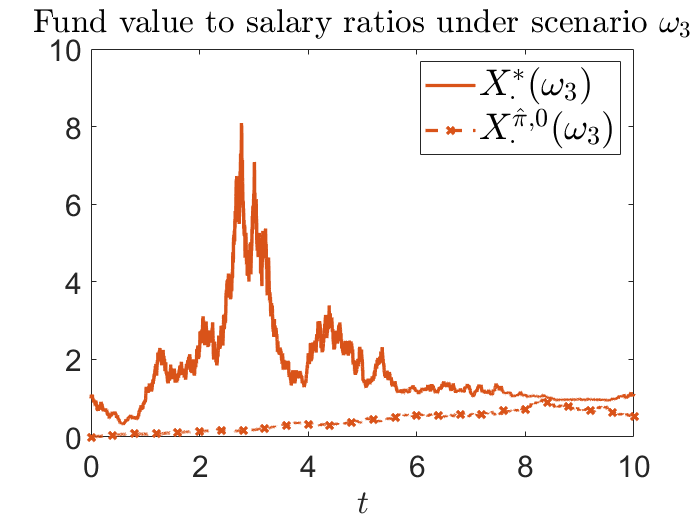}
\caption{$\beta\equiv -0.25$}
\label{fig:power:XZ:US:b-05}
\end{subfigure}%
\begin{subfigure}{.5\textwidth}
\centering
\includegraphics[scale=0.4]{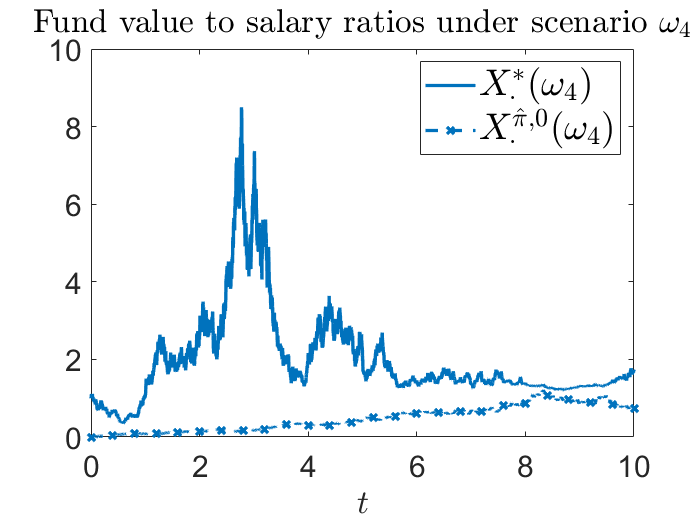}
\caption{$\beta\equiv -0.25$}
\label{fig:power:XZ:DS:b-05} 
\end{subfigure}%

\begin{subfigure}{.5\textwidth}
\centering
\includegraphics[scale=0.4]{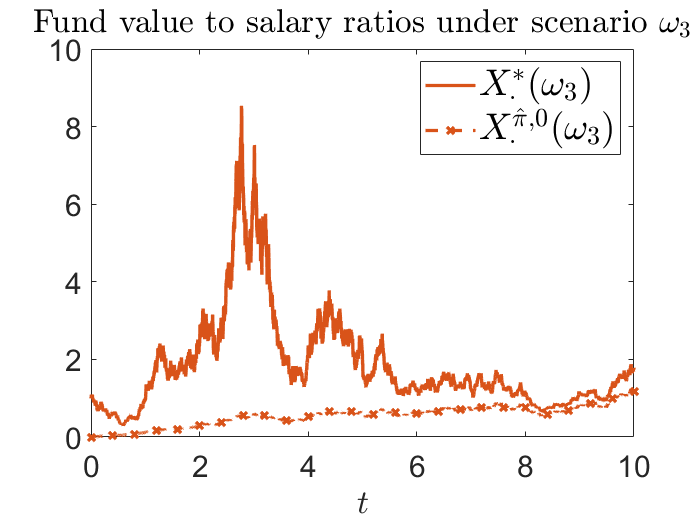}
\caption{$\beta\equiv 0.25$}
\label{fig:power:XZ:US:b05}
\end{subfigure}%
\begin{subfigure}{.5\textwidth}
\centering
\includegraphics[scale=0.4]{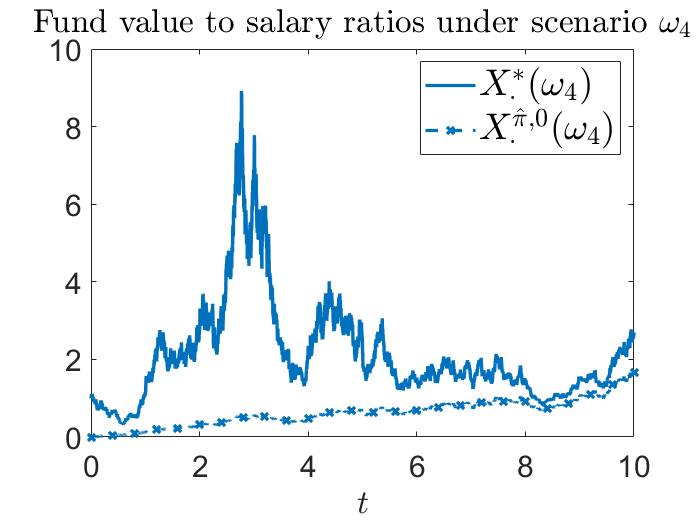}
\caption{$\beta\equiv 0.25$}
\label{fig:power:XZ:DS:b05} 
\end{subfigure}%

\caption{Realizations of worker's pension fund value to salary ratios under scenarios $\omega_3$ and $\omega_4$ and with respect to two different baseline performances}
\label{fig:power:XZ:S}
\end{figure}

Similar interpretation, in terms of the worker's optimal investment strategy being a convex combination of the baseline and myopic strategies, is also clearly shown in Figure \ref{fig:power:Pi:S}; see also Figure \ref{fig:power:XZ:S} for the corrseponding fund value to salary ratios, of optimality and of baseline. However, note that her optimal investment strategies, when the non-hedgeable risk $B^2$ moves up under the scenario $\omega_3$, and when $B^2$ moves down under the scenario $\omega_4$, are almost identical. This is because the diffusion of the fund value to salary ratio with respect to the non-hedgeable risk $B^2$ is independent of her investment strategy; the influence on the ratio by different $B^2$ is not amplified by non-identical investment strategies.

{\color{black}\subsection{Motivating Example Revisited}\label{sec:mot_eg_revisited}


Recall that, in Section \ref{sec:backward:pitfall}, by considering that the worker is promoted at time $t_0\in\left(0,T\right)$ which is not anticipated when she plans her investment strategy at time $t=0$, the motivating example showed the incompatibility of optimality and time-consistency of an environment-adapting investment strategy under the classical backward model and approach. In this section, we revisit this motivating example and demonstrate the compatibility of optimality and time-consistency of an environment-adapting investment strategy counterpart under the forward model and approach.\\




At the current time $t=0$, the setting reconciles that in Section \ref{sec:backward:pitfall}, except that the worker evaluates her future pension fund value to salary ratio by a forward preference. Suppose that her forward preference is given by the power forward utility preference in \eqref{eq:U:power}. Therefore, by Theorem \ref{pp:power}, her optimal investment strategy $\pi^{\text{for}}_{\left[0,\infty\right)}=\{ \pi^{\text{for}}_{t,\left[0,\infty\right)}\}_{t\geq 0}$, which is solved by the forward approach, given by, for any $t\geq 0$,
\begin{equation}
\pi^{\text{for}}_{t,\left[0,\infty\right)}:=\frac{\sigma^{Y,1}}{\sigma}+\frac{\lambda-\sigma^{Y,1}+\theta^1}{\sigma\left(1-\gamma\right)}\left(1+\left(\frac{\beta\left(1-\gamma\right)}{\lambda-\sigma^{Y,1}+\theta^1}-1\right)\frac{X^{\hat{\pi},0}_t}{X_t}\right),
\label{eq:pi_forward}
\end{equation}
solves the following time-$0$ planning problem:
\begin{equation}
\sup_{\pi\in\mathcal{A}}\mathbb{E}\left[U\left(X^{\pi}_T,T\right)\right],
\label{eq:V:forward:0}
\end{equation}
where $X^{\hat{\pi},0}=\left\{X^{\hat{\pi},0}_t\right\}_{t\geq 0}$ solves, for any $t\geq 0$,
\begin{equation*}
dX^{\hat{\pi},0}_t = pdt + X^{\hat{\pi},0}_t\left( \left(\sigma^{Y,1}\lambda+\beta\left(\lambda - \sigma^{Y,1}\right) - \mu^Y  \right)dt +\beta dB^1_t \right),
\end{equation*}
with $X^{\hat{\pi},0}_0=0$. Note that her investment strategy $\pi^{\text{for}}_{t,\left[0,\infty\right)}$ at any time $t\geq 0$, given in \eqref{eq:pi_forward}, depends only on the assumed model dynamics on the time horizon $\left[0,t\right]$, but not on that from the time $t$ onward; this contrasts with the counterpart $\pi^{\text{back}}_{t,\left[0,T\right]}$ given in \eqref{eq:pi:back}, which depends also on the assumed model dynamics over the future time horizon $\left[t,T\right]$, due to the backward model and approach.\\

Suppose that the worker {\color{black}forward} implemented the investment strategy $\pi^{\text{for}}_{\left[0,t_0\right)}$ up to the time $t_0\in\left(0,T\right)$, when she is promoted and the risk premium of her salary increases from $\mu^Y$ to $\tilde{\mu}^{Y}$, and ceteris paribus. At the current time $t=t_0$, with the unexpected change being revealed, the worker could either:
\begin{enumerate}
\item[(i)] adapt to the changed environment, and {\color{black}forward} implement the investment strategy $\pi^{\text{for},1}_{\left[t_0,\infty\right)}=\{ \pi^{\text{for},1}_{t,\left[t_0,\infty\right)}\}_{t\geq t_0}$ from the time $t_0$ onward of the following time-$t_0$ planning problem, solved by the forward approach:
\begin{equation}
\sup_{\pi\in\mathcal{A}}\mathbb{E}_{t_0}\left[U\left(X^{\pi}_T,T\right)\right];
\label{eq:v:forward:t0}
\end{equation}
her {\color{black}forward} implementing investment strategy from time $t=0$ onward would be given by $\pi^{\text{for},1}_{\left[0,\infty\right)}=\pi^{\text{for}}_{\left[0,t_0\right)}\oplus\pi^{\text{for},1}_{\left[t_0,\infty\right)}$, which is, for any $t\geq 0$,
\begin{equation}
\pi^{\text{for},1}_{t,\left[0,\infty\right)}:=\frac{\sigma^{Y,1}}{\sigma}+\frac{\lambda-\sigma^{Y,1}+\theta^1}{\sigma\left(1-\gamma\right)}\left(1+\left(\frac{\beta\left(1-\gamma\right)}{\lambda-\sigma^{Y,1}+\theta^1}-1\right)\frac{\tilde{X}^{\hat{\pi},0}_t}{X_t}\right),
\label{eq:pi:forward_1}
\end{equation}
where $\tilde{X}^{\hat{\pi},0}=\left\{\tilde{X}^{\hat{\pi},0}_t\right\}_{t\geq 0}$ solves, for any $t\geq 0$,
\begin{align*}
d\tilde{X}^{\hat{\pi},0}_t =&\;pdt\\&+ \tilde{X}^{\hat{\pi},0}_t\left( \left(\sigma^{Y,1}\lambda+\beta\left(\lambda - \sigma^{Y,1}\right) - \left(\mu^Y\mathds{1}_{\left[0,t_0\right)}\left(t\right)+\tilde{\mu}^Y\mathds{1}_{\left[t_0,\infty\right)}\left(t\right)\right)  \right)dt +\beta dB^1_t \right),
\end{align*}
with $X^{\hat{\pi},0}_0=0$;
\item[(ii)] commit to the pre-specified environment on solving the time-$0$ planning problem \eqref{eq:V:forward:0}, and {\color{black}forward} implement the investment strategy $\pi_{\left[t_0,\infty\right)}^{\text{for},2}=\{ \pi^{\text{for},2}_{t,\left[t_0,\infty\right)}\}_{t\geq t_0}$ from the time $t_0$ onward as the same as the optimal investment strategy $\pi_{\left[t_0,\infty\right)}^{\text{for}}$ of the time-$0$ planning problem \eqref{eq:V:forward:0}; her {\color{black}forward} implementing investment strategy from time $t=0$ onward would be given by $\pi_{\left[0,\infty\right)}^{\text{for},2}=\pi_{\left[0,t_0\right)}^{\text{for}}\oplus\pi_{\left[t_0,\infty\right)}^{\text{for}}=\pi_{\left[0,\infty\right)}^{\text{for}}$, which is given in \eqref{eq:pi_forward}.
\end{enumerate}

By definition, the investment strategies $\pi_{\left[t_0,\infty\right)}^{\text{for},1}$ and $\pi_{\left[t_0,\infty\right)}^{\text{for},2}$ are, respectively, optimal and sub-optimal to the time-$t_0$ planning problem \eqref{eq:v:forward:t0}. Most importantly, the inevitable inequality that $\pi_{\left[t_0,\infty\right)}^{\text{for},1}\neq\pi_{\left[t_0,\infty\right)}^{\text{for}}$, because of the adaptive nature of the optimal investment strategy $\pi_{\left[t_0,\infty\right)}^{\text{for},1}$, does not imply any time-inconsistent issues. Indeed, if the worker had prophesied the exact change in the environment, she would have {\color{black}forward} implemented the optimal investment strategy $\pi^{\text{for},*}_{\left[0,\infty\right)}=\{ \pi^{\text{for},*}_{t,\left[0,\infty\right)}\}_{t\geq 0}$ of the time-$0$ planning problem \eqref{eq:V:forward:0} from time $t=0$ onward, which turns out to be equivalent to the environment-adapting investment strategy $\pi^{\text{for},1}_{\left[0,\infty\right)}$, due to the forward model and approach. The equality that $\pi_{\left[0,t_0\right)}^{\text{for},*}=\pi_{\left[0,t_0\right)}^{\text{for}}$ entails that the worker would make the same optimal investment decision to the time-$0$ planning problem \eqref{eq:V:forward:0} at any time before the change of environment is unfolded, whether or not the knowledge of this development is a priori known at the time $t=0$; in other words, at the current time $t=t_0$, the worker does not regret her implemented investment strategy $\pi_{\left[0,t_0\right)}^{\text{for}}=\pi_{\left[0,t_0\right)}^{\text{for},1}=\pi_{\left[0,t_0\right)}^{\text{for},2}=\pi_{\left[0,t_0\right)}^{\text{for},*}$, even though $\pi_{\left[t_0,\infty\right)}^{\text{for},1}\neq\pi_{\left[t_0,\infty\right)}^{\text{for}}$. This vigorously contrasts with the backward counterpart being discussed in Section \ref{sec:backward:pitfall}, and highlights that the forward model and approach allows the environment-adapting investment strategy to maintain optimality and time-consistency at once.\\

Since the environment-adapting investment strategy $\pi_{\left[0,\infty\right)}^{\text{for},1}$ is shown to be desirable, this section concludes by comparing it with the sub-optimal pre-commitment strategy $\pi_{\left[0,\infty\right)}^{\text{for},2}$. Consider the same set of constant parameters as in Sections \ref{sec:backward:pitfall} and \ref{sec:power:numeric}, with $\beta=0.25$. Figure \ref{fig:forward:cdf} displays the empirical CDF of the difference $\pi_{15,\left[0,\infty\right)}^{\text{for},1}-\pi_{15,\left[0,\infty\right)}^{\text{for},2}$ at time $t=15$ after the worker is promoted. It shows that the worker would always over-invest in the risky asset if she commits to the pre-specified environment.\\

    \begin{figure}[!h]
        \centering
        \includegraphics[scale=0.5]{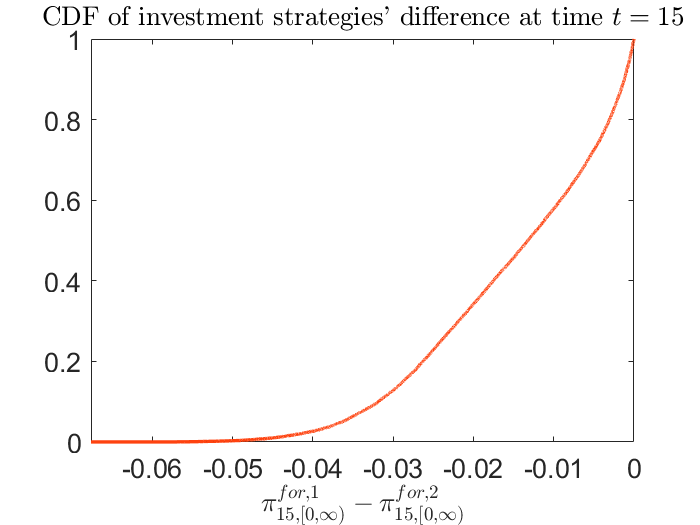}
        \caption{\color{black}Empirical cumulative distribution function of investment strategies' difference $\pi_{t,\left[0,\infty\right)}^{\text{for},1}-\pi_{t,\left[0,\infty\right)}^{\text{for},2}$, at time $t=15$, after the change of environment}
        \label{fig:forward:cdf}
    \end{figure}

}

\section{Exponential Forward Utility Preferences}
\label{sec:exp}
This section constructs an exponential forward utility preference on the worker's pension fund value to salary ratio. Unlike the constructed power forward utility preferences in Section \ref{sec:power}, the stochastic domain $\mathcal{D}\equiv\mathbb{R}$ in this case, as $\inf\mathcal{D}_0=-\infty$; also, the process $Z\equiv 0$, and thus the exponential forward preference is the translated random field itself with $\tilde{\mathcal{D}}=\mathbb{R}$. In the following, though, the same baseline investment strategy shall still serve a critical role.\\

Let $\Gamma:=\{\Gamma_t\}_{t\geq 0}\in\mathcal{P}_1(\mathbb{F})$ be a process, such that (i) it satisfies that $\Gamma_0^{-1}=\gamma>0$, which is the worker's risk aversion parameter at the current time $t=0$, and (ii) it admits the following It\^o's diffusion form: for any $t\geq 0$,
\begin{equation*}
\label{eq:Gamma:exp}
d\Gamma_t = \Gamma_t\left(\alpha_tdt + \beta_t^\top d{\bf B}^1_t -(\sigma^{Y,2}_t)^\top d{\bf B}^2_t \right),
\end{equation*}
where $\alpha=\{\alpha_t\}_{t\geq 0}\in\mathcal{P}_1\left(\mathbb{F}\right)$ and $\beta=\{\beta_t\}_{t\geq 0}\in\mathcal{P}_n\left(\mathbb{F}\right)$ are uniformly bounded and also satisfy the same relation \eqref{eq:power:alpha:beta}{\color{black}. T}he dynamics of $\Gamma$ can {\color{black}thus} be rewritten as, for any $t\geq 0$,
    \begin{align*}
    d\Gamma_t =&\;\Gamma_t\left( \left( \hat{\pi}^\top_t\Sigma_t (\lambda_t-\sigma^{Y,1}_t) - \mu^Y_t + \|\sigma^{Y,1}_t\|^2 + \|\sigma^{Y,2}_t\|^2 \right) dt  \right.\\&\quad\quad\left.+ (\hat{\pi}_t^\top\Sigma_t-(\sigma^{Y,1}_t)^\top) d{\bf B}^1_t - (\sigma^{Y,2}_t)^\top d{\bf B}^2_t \right),
\end{align*}
where $\hat{\pi}_t:=(\Sigma_t^\top)^{-1}(\sigma^{Y,1}_t+\beta_t)$, for $t\geq 0$, and thus $\Gamma$ is also a fund value to salary ratio, under the same exogenous baseline investment strategy $\hat{\pi}$, but with zero contribution from the worker as well as the ratio starting at $\gamma^{-1}$. Assigning the reciprocal of the worker's current risk aversion as the initial baseline fund value to salary ratio could seem to be artificial at the first glance; yet, the constructed exponential forward utility preference below shall designate another economic interpretation to $\Gamma$, as its reciprocal $\Gamma^{-1}:=\{\Gamma^{-1}_t\}_{t\geq 0}\in\mathcal{P}_1(\mathbb{F})$ is a dynamic risk aversion process of the worker.


\subsection{Admissibility}\label{sec:admis_exponential}
To define the admissible set of investment strategies in this case, only an integrability condition for constructing the preference is imposed:
\begin{equation*}
\label{eq:admissible:exp}
\mathcal{A}= \left\{ \pi \in \mathcal{P}_n(\mathbb{F}) : \frac{X^\pi}{\Gamma}(\pi- \hat{\pi}) \in \mathcal{L}^2_{n,\text{BMO}}  \right\}.
\end{equation*}
Loosely speaking, an investment strategy is admissible if it does not deviate too much from the baseline strategy, in terms of the $\mathcal{L}^2_{n,\text{BMO}}$-norm, with the deviation weighed by the quotient of the two ratios $X^{\pi}$ and $\Gamma$. {\color{black}Since the risk aversion $\Gamma^{-1}$ herein is stochastic, it is naturally taken into consideration for the integrability with respect to the strategy; see also the condition in Theorem 3.5 of \cite{michail:exponential:2014}. For a more direct justification of the integrability condition in $\mathcal{A}$ by the relevant literature, see Section \ref{section:exp_W} below.} The following equivalent form of the admissible set shall be more handy:
    \begin{equation*}
        \label{eq:A:tile:exp}
        \tilde{\mathcal{A}} = \left\{ \pi \in \mathcal{P}_n(\mathbb{F}) : \pi_t  = \hat{\pi}_t + \frac{\Gamma_t}{X^\pi_t}\xi_t,\;t\geq 0,\text{ for some }\xi = \{\xi_t\}_{t\geq 0}\in \mathcal{L}^2_{n,\text{BMO}} \right\}.
    \end{equation*}
\begin{proposition}
\label{pp:admissible:exp}
We have $\mathcal{\Tilde{A}}=\mathcal{A}$.
\end{proposition}

\begin{proof}
Let $\pi\in\mathcal{\Tilde{A}}$. For any $t\geq 0$, $\frac{X^\pi_t}{\Gamma_t}(\pi_t- \hat{\pi}_t)=\xi_t$. Therefore, $\frac{X^\pi}{\Gamma}(\pi- \hat{\pi})\equiv\xi \in \mathcal{L}^2_{n,\text{BMO}}$, and thus $\pi\in\mathcal{A}$. This shows $\mathcal{\Tilde{A}}\subseteq \mathcal{A}$. The other inclusion $\mathcal{A}\subseteq\mathcal{\Tilde{A}}$ is clear.
\end{proof}

\subsection{Non-Zero Volatility Forward Preferences}\label{sec:exp:main}
The following theorem constructs a (non-)zero volatility exponential forward utility preference of the worker, together with her corresponding optimal investment strategy. Their economic insights shall be highlighted after the theorem{\color{black}; its proof is relegated to Appendix \ref{sec:app:proof:exp}.}
\begin{theorem}
\label{pp:exp}
Let $V=\left\{V_t\right\}_{t\geq 0}\in\mathcal{P}_1\left(\mathbb{F}\right)$ be a process, given by $V_0=0$ and, for any $t\geq 0$,
\begin{equation*}
dV_t=\frac{1}{2}\left(\|\lambda_t-\sigma^{Y,1}_t+\theta^1_t-\beta_t\|^2-\|\theta^1_t\|^2-\|\theta^2_t\|^2\right)dt +(\theta^1_t)^\top d{\bf B}^1_t + (\theta^2_t)^\top d{\bf B}^2_t,
\end{equation*}
where $\theta^1 = \{\theta^1_t\}_{t\geq 0}\in \mathcal{P}_n(\mathbb{F})$ and $\theta^2 =\{\theta^2_t\}_{t\geq 0} \in \mathcal{P}_m(\mathbb{F})$ are uniformly bounded. The random field, for any $t\geq 0$ and $x\in\mathbb{R}$,
\begin{equation}
\label{eq:U:exp}
U\left(x,t\right)=-\exp\left( - \frac{x-X^{\hat{\pi},0}_t}{\Gamma_t} + V_t \right),
\end{equation}
where $X^{\hat{\pi},0}:=\left\{X^{\hat{\pi},0}_t\right\}_{t\geq 0}$ solves \eqref{eq:Z:power:rewrite} with $X^{\hat{\pi},0}_0=0$, is an exponential forward utility preference on the fund value to salary ratio. In addition, its volatility processes are given by, for any $x\in\mathbb{R}$ and $t\geq 0$,
\begin{equation}
\label{eq:vol:exp}
\begin{aligned}
      a_1(x,t) =&-\exp\left(-\frac{x-X^{\hat{\pi},0}_t}{\Gamma_t}+V_t \right)\left(\theta^1_t+\frac{x}{\Gamma_t}\beta_t \right),\text{ and} \\
      a_2(x,t) =&-\exp\left(-\frac{x-X^{\hat{\pi},0}_t}{\Gamma_t}+V_t \right)\left(\theta^2_t-\frac{x}{\Gamma_t} \sigma^{Y,2}_t\right).
\end{aligned}
\end{equation}
Moreover, the optimal investment strategy is given by, for any $t\geq 0$,
\begin{equation}
\label{eq:pi*:exp}
\pi_t^* = \frac{\Gamma_t}{X^*_t}\left(\Sigma^{\top}_t\right)^{-1}\left(\lambda_t+\theta^1_t\right)+\left(1-\frac{\Gamma_t}{X^*_t}\right)\hat{\pi}_t,
\end{equation}
where $X^*:\equiv X^{\pi^*}$ satisfies, for any $t\geq 0$,
\begin{align*}
dX_t^* =& \  \Big(p_t + \Gamma_t(\lambda_t-\sigma^{Y,1}_t)^\top(\lambda_t-\sigma^{Y,1}_t+\theta^1_t-\beta_t)   \\ & \quad +  X^*_t\left( (\lambda_t-\sigma^{Y,1}_t)^\top\beta_t+\lambda^\top_t\sigma^{Y,1}_t + \|\sigma^{Y,2}_t\|^2 - \mu^Y_t \right)\Big)dt   \\ & \
+   \left(\Gamma_t(\lambda_t-\sigma^{Y,1}_t+ \theta^1_t-\beta_t)+X^*_t \beta_t  \right)^\top d{\bf B}^1_t   - X^*_t(\sigma^{Y,2}_t)^\top d{\bf B}_t^2 .
\end{align*}



\end{theorem}
The worker's forward utility preference depends on the two baseline fund value to salary ratios in this case. One is $X^{\hat{\pi},0}$ with the salary contribution and starting from zero fund value, while another one is $\Gamma$ with zero salary contribution and the initial ratio as the reciprocal of the worker’s current risk aversion. Again, the worker determines how much more her fund value to salary ratio exceeding the baseline ratio $X^{\hat{\pi},0}$; however, unlike the power forward utility preference therein, this surplus $x-X^{\hat{\pi},0}$ could be negative, that her ratio could be less than the baseline ratio. The surplus first measures the worker's fund performance comparing to the baseline, in terms of the ratio, without taking the explicit effect by the salary contribution into account. This absolute difference is then contrasted to the other baseline ratio $\Gamma$, which is without any explicit or implicit considerations of the salary contribution. Therefore, $\frac{x-X^{\hat{\pi},0}}{\Gamma}$ quantifies the relative difference of the worker's fund performance, with respect to the baseline investment strategy.\\

Recall that $U\left(x,0\right)=-\exp\left(-\gamma x\right)=-\exp\left(-\Gamma^{-1}_0 x\right)$, for $x\in\mathbb{R}$, and thus the reciprocal of the initial baseline ratio $\Gamma^{-1}_0$ measures the worker's current risk aversion, by definition. This relationship can be generalized to any future times. That is, define the process $\Gamma^{-1}=\{\Gamma^{-1}_t\}_{t\geq 0}$ as the reciprocal of the baseline ratio; then, for any $t\geq 0$, $U\left(x,t\right)=-\exp\left(-\Gamma^{-1}_t\left(x-X^{\hat{\pi},0}_t\right)+ V_t \right)$, which resembles standard exponential preferences with constant risk aversion parameter in the place of $\Gamma^{-1}$, and thus $\Gamma^{-1}$ can be perceived as the dynamic risk aversion process of the worker. In such manner, if the baseline ratio $\Gamma$ performs well with a large value, the worker will be less risk averse, and vice versa; this is consistent with the fact that any admissible investment strategies are not deviating too much from the baseline strategy.\\


The worker’s optimal investment strategy in \eqref{eq:pi*:exp}, based on her exponential forward utility preference, is a combination of the baseline strategy $\hat{\pi}$ and a myopic strategy $\left(\Sigma^{\top}_t\right)^{-1}\left(\lambda_t+\theta^1_t\right)$, for $t\geq 0$. In the case of $\theta^1\equiv 0$, this myopic component resembles the optimal investment strategy under the Merton's classical investment problem. {\color{black} Due to the forward nature of the preference, the optimal investment strategy at any time is again independent of the model of the environment in the future, making model updates possible, if any.} Different from the optimal investment strategy in \eqref{eq:pi*:power:1} derived from power forward utility preference, the optimal investment strategy in \eqref{eq:pi*:exp} is not necessarily being confined between the baseline strategy and the myopic strategy. \\

A numerical example for exponential preferences shall be omitted, as the illustration of the stochasticity of the preference and the worker's optimal investment strategy are similar to that in Section \ref{sec:power:numeric}. {\color{black}If the motivating example is revisited as in Section \ref{sec:mot_eg_revisited}, the same conclusion also holds that the environment-adapting investment strategy using the exponential forward utility preferences is both optimal and time-consistent.}

{\color{black}\section{Absolute Fund Value as Quantity of Interest}\label{sec:new_section}
Before parting by the concluding remarks and future directions, it is worth to reiterate that the consideration of the worker's pension fund value to salary ratio $X$ as her quantity of interest throughout this paper is due to the habit-formation in life-cycle models. This advocate is not because of any mathematical tractability. Even if her absolute pension fund value $W$, with its dynamics given in \eqref{eq:fund_value_dynamics}, is set to be evaluated, a CRRA power or CARA exponential forward utility preference on it can be constructed in a similar manner as in Sections \ref{sec:power} or \ref{sec:exp}, by considering an \textit{ansatz} which compares with the absolute fund value under an exogenous baseline investment strategy.

To this end, let $\tilde{\pi}=\{\tilde{\pi}_t\}_{t\geq 0}\in \mathcal{P}_n(\mathbb{F})$ be the exogenous uniformly bounded baseline investment strategy, and let $W^{\tilde{\pi},0}=\left\{W^{\tilde{\pi},0}_t\right\}_{t\geq 0}$ be the absolute fund value satisfying the dynamics given in \eqref{eq:fund_value_dynamics}, with implementing the baseline strategy $\tilde{\pi}$, and with the zero initial value, i.e., $W^{\tilde{\pi},0}_0=0$.

\subsection{Power Forward Utility Preferences}
Define the admissible set of investment strategies by
\begin{equation*}
\mathcal{A}^W:= \left\{ \pi \in\mathcal{P}_n(\mathbb{F}): W^\pi \pi \in \mathcal{L}^2_n, \ W^\pi_t > W^{\tilde{\pi},0}_t, \ \text{for a.a. } (t,\omega) \in [0,\infty)\times\Omega \right\}.
\end{equation*}
Let $\tilde{V}=\{\tilde{V}_t\}_{t\geq 0}\in\mathcal{P}_1(\mathbb{F})$ be a process, given by $\tilde{V}_0=0$, and, for any $t\geq 0$,
\begin{equation*}
d\tilde{V}_t = \tilde{v}_t dt + (\tilde{\theta}^1_t)^\top d{\bf B}^1_t + (\tilde{\theta}^2_t)^\top d{\bf B}^2_t,
\end{equation*}
where $\tilde{v}=\{\tilde{v}_t\}_{t\geq 0}\in \mathcal{P}_1(\mathbb{F})$, $\tilde{\theta}^1 = \{\tilde{\theta}^1_t\}_{t\geq 0}\in \mathcal{P}_n(\mathbb{F})$, and $\tilde{\theta}^2 = \{\tilde{\theta}^2_t\}_{t\geq 0}\in \mathcal{P}_m(\mathbb{F})$ are uniformly bounded processes, such that, for any $t\geq 0$,
\begin{equation*}
\tilde{v}_t = -r\gamma - \frac{\gamma\|\lambda_t +  \tilde{\theta}^1_t\|^2}{2(1-\gamma) } - \frac{\|\tilde{\theta}^1_t\|^2+ \|\tilde{\theta}^2_t\|^2}{2}.
\end{equation*}
Define a random field, for any $t\geq 0$ and $w\in\mathbb{R}$,
\begin{equation}
U^W\left(w,t\right) := \begin{cases}
\frac{1}{\gamma}\left(w-W^{\tilde{\pi},0}_t\right)^\gamma e^{\tilde{V}_t}, &\text{if } w >W^{\tilde{\pi},0}_t;\\
-\infty, &\text{otherwise}.
\end{cases}
\label{eq:U_W_power}
\end{equation}
Following the same proof of Theorem \ref{pp:power}, the random field is a power forward utility preference on the absolute pension fund value, and the optimal investment strategy is given by, for any $t\geq 0$,
\begin{equation}
\pi^{W,*}_t = \frac{W^{\tilde{\pi},0}_t}{W^*_t} \tilde{\pi}_t + \left(1-\frac{W^{\tilde{\pi},0}_t}{W^*_t} \right) \left(\Sigma_t^\top\right)^{-1}\frac{\lambda_t + \tilde{\theta}^1_t}{1-\gamma},
\label{eq:U_W_power_strategy}
\end{equation}
where $W^*=\left\{W^*_t\right\}_{t\geq 0}$ is the absolute pension fund value process implementing the optimal strategy $\pi^{W,*}$ in \eqref{eq:fund_value_dynamics}.

\subsection{Exponential Forward Utility Preferences}\label{section:exp_W}
Let $\tilde{\Gamma}:=\left\{\tilde{\Gamma}_t\right\}_{t\geq 0}\in\mathcal{P}_1\left(\mathbb{F}\right)$ be a process which satisfies that $\tilde{\Gamma}^{-1}_0=\gamma$, and, for any $t\geq 0$,
\begin{equation*}
d\tilde{\Gamma}_t = \tilde{\Gamma}_t\left(\left(r+\tilde{\pi}_t^\top\mu_t\right)dt+\tilde{\pi}^\top_t\Sigma_td{\bf B}^1_t\right);
\end{equation*}
that is, $\tilde{\Gamma}$ is an absolute pension fund value, under the exogenous baseline strategy $\tilde{\pi}$, but with zero contribution from the worker as well as the value starting at $\gamma^{-1}$. In turn, define the admissible set of investment strategies by
\begin{equation*}
\mathcal{A}^W := \left\{\pi \in\mathcal{P}_n(\mathbb{F}):\frac{W^\pi}{\tilde{\Gamma}}(\pi-\tilde{\pi})\in \mathcal{L}^2_{n,\text{BMO}}   \right\}.
\end{equation*}
This integrability condition is in line with the construction of forward utility preferences in the literature, where the worker's risk aversion is a constant and she does not compare her investment strategy to an exogenous baseline strategy. {\color{black} In that case, we have $\tilde{\pi} = 0$ and the process $\tilde{\Gamma}$ is a constant. The admissible set is then reduced to requiring $W^\pi \pi \in \mathcal{L}^2_{n,\text{BMO}}$}; see, for example, {\color{black}Section 4 of \cite{liang:bsde} and Section 2.5 of \cite{CHONG201993}}.

Let $\tilde{V}=\{\tilde{V}_t\}_{t\geq 0}\in\mathcal{P}_1(\mathbb{F})$ be a process, given by $\tilde{V}_0=0$, and, for any $t\geq 0$,
\begin{equation*}
d\tilde{V}_t = \frac{1}{2}\left(\|\lambda_t + \tilde{\theta}^1_t - \Sigma_t^\top\tilde{\pi}_t\|^2 -\|\tilde{\theta}^1_t\|^2-\|\tilde{\theta}^2_t\|^2 \right)dt + (\tilde{\theta}^1_t)^\top d{\bf B}^1_t + \tilde{\theta}_t^2d{\bf B}^2_t. 
\end{equation*}
where $\tilde{\theta}^1 = \{\tilde{\theta}^1_t\}_{t\geq 0}\in \mathcal{P}_n(\mathbb{F})$ and $\tilde{\theta}^2 = \{\tilde{\theta}^2_t\}_{t\geq 0}\in \mathcal{P}_m(\mathbb{F})$ are uniformly bounded. Define a random field, for any $t\geq 0$ and $w\in\mathbb{R}$,
\begin{equation}
U^W\left(w,t\right) := -\exp\left( -\frac{w-W^{\tilde{\pi},0}_t}{\tilde{\Gamma}_t} + \tilde{V}_t \right).
\label{eq:U_W_exponential}
\end{equation}
Following the same proof of Theorem \ref{pp:exp}, the random field is an exponential forward utility preference on the absolute pension fund value, and the optimal investment strategy is given by, for any $t\geq 0$,
\begin{equation}
\pi^{W,*}_t = \frac{\tilde{\Gamma}_t}{W^*_t} \left(\Sigma_t^\top\right)^{-1}  \left(\lambda_t+\tilde{\theta}^1_t\right) + \left(1 - \frac{\tilde{\Gamma}_t}{W^*_t} \right) \tilde{\pi}_t.
\label{eq:U_W_exponential_strategy}
\end{equation}

\subsection{Comparison to Ratio as Quantity of Interest}
In both cases, it is evident that the coefficients $\mu^{Y}_{\cdot}$, $\sigma^{Y,1}_{\cdot}$, and $\sigma^{Y,2}_{\cdot}$ do not contribute directly to the forward utility preferences, in \eqref{eq:U_W_power} and \eqref{eq:U_W_exponential}, and their corresponding optimal investment strategies, in \eqref{eq:U_W_power_strategy} and \eqref{eq:U_W_exponential_strategy}, comparing with the counterparts in Sections \ref{sec:power} and \ref{sec:exp} using the pension fund value to salary ratio as the quantity of interest for the worker. Her realized salary, though, would certainly affect the evolution of her absolute fund value $W$, as well as the baseline one $W^{\tilde{\pi},0}$, and consequently influence her realized forward preferences and optimal investment strategies.}

\section{Concluding Remarks and Future Directions}
\label{sec:conclusion}
This paper solved the optimal investment strategy of the worker, who is enrolled to the DC pension scheme, during the accumulation phase. Instead of following the mainstream in the literature using the backward model and approach, this paper proposed to decide her optimal strategy based on the forward model and approach, via her forward utility preferences. Such a forward methodology allowed the worker to, flexibly decide her actual retirement time in the future, and derive her future preferences due to the actual realizations in the market environment. This paper discussed the SPDE representation for the worker's forward preferences, featuring their non-uniqueness and volatility processes. Two of her forward utility preferences were then constructed, with the corresponding optimal investment strategies being solved, respectively in the cases of power and exponential utility functions as her current preference. In both cases, her forward utility preferences were based on the comparison of her fund value to salary ratio to another ratio generated by the exogenous baseline investment strategy. In the power utility case, the worker's optimal investment strategy generalized the CPPI strategy with the stochastic floor, the stochastic multiple of the cushion, and the additive factor. In the exponential case, the reciprocal of the similar exogenous ratio, without the consideration of salary contribution, was shown to be the dynamic risk aversion process of the worker.\\

{\color{black} In the classical backward model and approach, in order to obtain the optimal strategy, the model for the entire planning horizon has to be precisely specified initially. Consequently, any changes in the model during the planning horizon will inevitably lead to the sub-optimality or time-inconsistency of the implemented strategy. In contrast, under the forward model and approach, the optimal investment strategy only depends on the model to-date. Thus, a}dopting the forward utility preferences {\color{black} would} allow regular model-updates for future dynamics together with strategy-revisions satisfying the time-consistency. This is especially applicable to solving the optimal investment strategy of the worker in the accumulation phase of the DC pension plan, which is often with a long horizon, and shall be revisited {\color{black} in a more generic manner} as one of the future directions.\\

Another future direction derived from this paper is to adopt the forward model and approach to plan for both accumulation and decumulation phases of the DC pension scheme. Only the backward model and approach have been developed in these combined problems. Yet, they are all subject to the same shortcoming that any pre-commitments would not hold in an even longer horizon together with the decumulation phase. The forward model and approach shall provide consistent and streamlining treatments for the plannings in both accumulation and decumulation phases.

\nocite{dosreisL:mean:field}
\nocite{BLAKE2014105}
\nocite{Nadtochiy2014}
\bibliographystyle{apacite}
\bibliography{ref}

{\color{black}\begin{appendices}
\section{Derivations of \eqref{eq:SPDE} and \eqref{eq:SPDE_strategy}}
\label{sec:app:SPDE}

For any $\pi\in\mathcal{A}$, by the It\^o-Wentzell formula, for any $t\geq 0$,
\begin{equation}
\begin{aligned}
&\;d\tilde{U}(\tilde{X}_t,t)\\=&\;b(\tilde{X}_t,t) dt + a_1(\tilde{X}_t,t)^\top d{\bf B}^1_t + a_2(\tilde{X}_t,t)^\top d{\bf B}^2_t + \tilde{U}_{\tilde{x}}(\tilde{X}_t,t)d\tilde{X}_t  \\ & \ + \frac{1}{2} \tilde{U}_{\tilde{x}\tilde{x}}(\tilde{X}_t,t) d\langle \tilde{X}_{\cdot}\rangle_t +   \left( X_t(\pi_t^\top\Sigma_t-(\sigma^{Y,1}_t)^\top) - (\kappa^1_t)^\top\right) \nabla_{\tilde{x}}a_1(\tilde{X}_t,t)dt  \\ & \ - \left(X_t(\sigma^{Y,2}_t)^\top + (\kappa_t^2)^\top  \right)\nabla_{\tilde{x}}a_2(\tilde{X}_t,t)  dt\\=&\;\Big( b(\tilde{X}_t,t) + \tilde{U}_{\tilde{x}}(\tilde{X}_t,t)\left(p_t -\nu_t + X_t\left( \pi_t^\top\Sigma_t(\lambda_t-\sigma^{Y,1}_t) - \mu^Y_t +  \|\sigma^{Y,1}_t\|^2 + \|\sigma^{Y,2}_t\|^2 \right)\right)\\&\;\quad+ \frac{\tilde{U}_{\tilde{x}\tilde{x}}(\tilde{X}_t,t)}{2} \left(\| X_t(\Sigma^\top_t\pi_t-\sigma^{Y,1}_t)-\kappa_t^1\|^2 + \|X_t\sigma^{Y,2}_t + \kappa_t^2\|^2\right)  \\
& \;\quad +\left( X_t(\pi_t^\top\Sigma_t-(\sigma^{Y,1}_t)^\top) - (\kappa^1_t)^\top\right) \nabla_{\tilde{x}}a_1(\tilde{X}_t,t)  - \left(X_t(\sigma^{Y,2}_t)^\top + (\kappa_t^2)^\top  \right)\nabla_{\tilde{x}}a_2(\tilde{X}_t,t) \Big)  dt\\&\; + (a_1(\tilde{X}_t,t)^\top + \tilde{U}_{\tilde{x}}(\tilde{X}_t,t)X_t(\pi^\top_t \Sigma_t-(\sigma^{Y,1}_t)^\top ) -(\kappa_t^1)^\top  ) d{\bf B}^1_t\\&\;  +( a_2(\tilde{X}_t,t)^\top -\tilde{U}_{\tilde{x}}(\tilde{X}_t,t)X_t (\sigma^{Y,2}_t)^\top - (\kappa^2_t)^\top  ) d{\bf B}^2_t.
\end{aligned}
\label{eq:SPDE1}
\end{equation}


Notice that the drift term of \eqref{eq:SPDE1} can be written as, for any $t\geq 0$,
\begin{align*}
 & \frac{  \tilde{U}_{\tilde{x}\tilde{x}}(\tilde{X}_t,t) X_t^2}{2} \left\| \Sigma^\top_t \pi_t - \sigma^{Y,1}_t+ \frac{\nabla_{\tilde{x}} a_1(\tilde{X}_t,t)+ (\lambda_t-\sigma^{Y,1}_t)\tilde{U}_{\tilde{x}}(\tilde{X}_t,t) -\kappa^1_t \tilde{U}_{\tilde{x}\tilde{x}}(\tilde{X}_t,t) }{ X_t \tilde{U}_{\tilde{x}\tilde{x}}(\tilde{X}_t,t) } \right\|^2  + A_t,
\end{align*}
where, 
\begin{align*}
A_t :=&\; b(\tilde{X}_t,t) +  \frac{(\|X_t\sigma^{Y,1}_t + \kappa^1_t\|^2+\|X_t\sigma^{Y,2}_t+\kappa_t^2\|^2)\tilde{U}_{\tilde{x}\tilde{x}}(\tilde{X}_t,t)}{2} \\
& - \left( X_t(\sigma^{Y,1}_t)^\top + (\kappa^1_t)^\top \right) \nabla_{\tilde{x}}a_1(\tilde{X}_t,t) - \left(X_t(\sigma^{Y,2}_t)^\top + (\kappa^2_t)^\top \right)\nabla_{\tilde{x}}a_2(\tilde{X}_t,t)   \\ & +\tilde{U}_{\tilde{x}}(\tilde{X}_t,t) (p_t-\nu_t + X_t( \|\sigma^{Y,1}_t\|^2+\|\sigma^{Y,2}_t\|^2 - \mu^Y_t))   \\ & - \frac{  \tilde{U}_{\tilde{x}\tilde{x}}(\tilde{X}_t,t) X_t^2}{2} \left\|   \sigma^{Y,1}_t- \frac{\nabla_{\tilde{x}} a_1(\tilde{X}_t,t)+ (\lambda_t-\sigma^{Y,1}_t)\tilde{U}_{\tilde{x}}(\tilde{X}_t,t) -\kappa^1_t \tilde{U}_{\tilde{x}\tilde{x}}(\tilde{X}_t,t) }{ X_t \tilde{U}_{\tilde{x}\tilde{x}}(\tilde{X}_t,t) } \right\|^2.\nonumber
\end{align*}
Now, let the drift process of the random field $\tilde{U}$ by, for any $\tilde{x}\in\mathcal{\tilde{D}}$ and $t\geq 0$,
\begin{align*}
b(\tilde{x},t):=&\;-  \frac{(\|(\tilde{x}+Z_t)\sigma^{Y,1}_t + \kappa^1_t\|^2+\|(\tilde{x}+Z_t)\sigma^{Y,2}_t+\kappa_t^2\|^2)\tilde{U}_{\tilde{x}\tilde{x}}(\tilde{x},t)}{2}   \\  &\;+
\left( (\tilde{x}+Z_t)(\sigma^{Y,1}_t)^\top + (\kappa^1_t)^\top \right) \nabla_{\tilde{x}}a_1(\tilde{x},t) + \left((\tilde{x}+Z_t)(\sigma^{Y,2}_t)^\top + (\kappa^2_t)^\top \right)\nabla_{\tilde{x}}a_2(\tilde{x},t) \\ &\ 
-  \tilde{U}_{\tilde{x}}(\tilde{x},t) (p_t-\nu_t + (\tilde{x}+Z_t)( \|\sigma^{Y,1}_t\|^2+\|\sigma^{Y,2}_t\|^2 - \mu^Y_t))   \\ &\; +\frac{  \tilde{U}_{\tilde{x}\tilde{x}}(\tilde{x},t)(\tilde{x}+Z_t)^2}{2} \left\|   \sigma^{Y,1}_t- \frac{\nabla_{\tilde{x}} a_1(\tilde{x},t)+ (\lambda_t-\sigma^{Y,1}_t)\tilde{U}_{\tilde{x}}(\tilde{x},t) -\kappa^1_t \tilde{U}_{\tilde{x}\tilde{x}}(\tilde{x},t) }{ (\tilde{x}+Z_t) \tilde{U}_{\tilde{x}\tilde{x}}(\tilde{x},t) } \right\|^2\\
=&\;- \frac{ \|(\tilde{x}+Z_t)\sigma^{Y,2}_t+\kappa_t^2\|^2\tilde{U}_{\tilde{x}\tilde{x}}(\tilde{x},t)}{2}    + \left((\tilde{x}+Z_t)\sigma^{Y,2}_t  + \kappa_t^2 \right)^\top\nabla_{\tilde{x}}a_2(\tilde{x},t) \\ &\ 
-  \tilde{U}_{\tilde{x}}(\tilde{x},t) \left(p_t-\nu_t + (\lambda_t-\sigma^{Y,1}_t)^\top \kappa^1_t+ (\tilde{x}+Z_t)( \lambda_t^\top\sigma^{Y,1}_t+\|\sigma^{Y,2}_t\|^2 - \mu^Y_t)\right)   \\ &\; +\frac{ \left\|     \nabla_{\tilde{x}} a_1(\tilde{x},t)+ (\lambda_t-\sigma^{Y,1}_t)\tilde{U}_{\tilde{x}}(\tilde{x},t)    \right\|^2 }{2 \tilde{U}_{\tilde{x}\tilde{x}}(\tilde{x},t)}  ,
\end{align*}
which leads to $A_\cdot\equiv 0$, and thus the drift term of \eqref{eq:SPDE1} is simply, for any $t\geq 0$,
\begin{equation*}
\frac{  \tilde{U}_{\tilde{x}\tilde{x}}(\tilde{X}_t,t) X_t^2}{2} \left\| \Sigma^\top_t \pi_t - \sigma^{Y,1}_t+ \frac{\nabla_{\tilde{x}} a_1(\tilde{X}_t,t)+ (\lambda_t-\sigma^{Y,1}_t)\tilde{U}_{\tilde{x}}(\tilde{X}_t,t) -\kappa^1_t \tilde{U}_{\tilde{x}\tilde{x}}(\tilde{X}_t,t) }{ X_t \tilde{U}_{\tilde{x}\tilde{x}}(\tilde{X}_t,t) } \right\|^2.
\end{equation*}
By the strict concavity of the random field $\tilde{U}$ in $\tilde{x}$, the process $\{\tilde{U}(\tilde{X}_t,t)\}_{t\geq 0}$ is indeed an $\mathbb{F}$-super-martingale under some integrability conditions. Furthermore, with, for any $t\geq 0$, $\{\tilde{U}(\tilde{X}_t,t)\}_{t\geq 0}$ is an $\mathbb{F}$-martingale under some integrability conditions. Therefore, $\tilde{U}$ is a forward preference on $\tilde{X}$, and hence $U$ is a forward preference on $X$.\\

\section{Proof of Theorem \ref{pp:power}}
        \label{sec:app:proof:power}
First, notice that $\pi^*$ can be rewritten as, for any $t\geq 0$,
\begin{equation*}
\pi_t^* = \frac{X^{\hat{\pi},0}_t}{X^*_t} \hat{\pi}_t + \left( 1-\frac{X^{\hat{\pi},0}_t}{X^*_t}\right) (\Sigma^\top_t)^{-1}\left(\sigma^{Y,1}_t+\frac{\lambda_t-\sigma^{Y,1}_t + \theta^1_t}{1-\gamma}\right),
\end{equation*}
and thus, by the (uniform) boundedness and Proposition \ref{pp:admissible}, $\pi^*\in\Tilde{\mathcal{A}}\subseteq\mathcal{A}$. Next, it is clear that $U$ in \eqref{eq:U:power} satisfies (i) and (ii) in Definition \ref{def:forward}. For verifying (iii) of Definition \ref{def:forward}, it suffices to show that $\{U(X^\pi_t,t)\}_{t\geq 0}$ is an $\mathbb{F}$-super-martingale for any $\pi\in\mathcal{A}$,  and that $\{U(X^*_t,t)\}_{t\geq 0}$ is an $\mathbb{F}$-martingale.\\

For any $\pi\in\mathcal{A}$, define the processes $\Tilde{X}=\{\Tilde{X}_t:=X^\pi_t-X^{\hat{\pi},0}_t\}_{t\geq 0}$, and $\tilde{\xi}=\{\tilde{\xi}_t\}_{t\geq 0}$ which is given by, for any $t\geq 0$, 
\begin{equation*}
\tilde{\xi}_t := \frac{X^\pi_t(\Sigma^\top_t \pi_t-\sigma^{Y,1}_t)-X^{\hat{\pi},0}_t\beta_t }{\Tilde{X}_t}.
\end{equation*}   
Then, $\Tilde{X}_0=x_0$ and $\Tilde{X}$ satisfies, for any $t\geq 0$,
\begin{equation*}
\begin{aligned}
d\Tilde{X}_t = & \ \Tilde{X}_t \Big(  \left(  \tilde{\xi}^\top_t(\lambda_t-\sigma^{Y,1}_t) + \lambda^\top_t\sigma^{Y,1}_t+\|\sigma^{Y,2}_t\|^2- \mu^Y_t  \right)  dt  \\ & \ \quad\quad+ \tilde{\xi}_t^\top d{\bf B}^1_t  -  (\sigma^{Y,2}_t)^\top d{\bf B}^2_t\Big).
\end{aligned}
\end{equation*}
Also, define $R_t^{\pi}:=\frac{1}{\gamma} (X^\pi_t-X^{\hat{\pi},0}_t)^\gamma e^{V_t}=\frac{1}{\gamma} \Tilde{X}_t^\gamma e^{V_t}$ for $t\geq 0$.\\

By It\^o's lemma, we obtain, for any $t\geq 0$,
    \begin{align*}
         \frac{dR_t^{\pi}}{R_t^{\pi}}  = & \ \frac{\gamma}{\tilde{X}_t} d\Tilde{X}_t + dV_t + \frac{\gamma(\gamma-1)}{2\tilde{X}_t^2} d\langle \tilde{X}_\cdot \rangle_t + \frac{1}{2}d\langle V_\cdot \rangle_t + \frac{\gamma}{\tilde{X}_t} d\langle  \Tilde{X}_\cdot ,V_\cdot\rangle_t \\ 
         = & \ \bigg( \gamma\left(  \tilde{\xi}^\top_t(\lambda_t-\sigma^{Y,1}_t) + \lambda^\top_t\sigma^{Y,1}_t+\|\sigma^{Y,2}_t\|^2- \mu^Y_t \right) + v_t  - \frac{\gamma(1-\gamma)}{2}(\|\tilde{\xi}_t\|^2+\|\sigma^{Y,2}_t\|^2)  \\ 
         & \ \quad + \frac{\|\theta^1_t\|^2+\|\theta_t^2\|^2}{2} + \gamma\left(\tilde{\xi}_t^\top \theta^1_t - (\sigma^{Y,2}_t)^\top \theta^2_t \right) \bigg) dt \\
         & \ + \left( \gamma \tilde{\xi}_t + \theta^1_t \right)^\top d{\bf B}^1_t + \left( \theta^2_t - \gamma\sigma^{Y,2}_t \right)^\top d{\bf B}^2_t \\
        = &  \ \bigg( -\frac{\gamma(1-\gamma)}{2}\left\| \tilde{\xi}_t - \frac{\lambda_t-\sigma^{Y,1}_t + \theta^1_t}{1-\gamma} \right\|^2 + v_t - \gamma\left((\theta^1_t)^\top\sigma^{Y,1}_t + (\theta^2_t)^\top \sigma^{Y,2}_t+\mu^Y_t \right) \\ 
        & \ \quad +\frac{\gamma(1+\gamma)\left(\|\sigma^{Y,1}_t\|^2 + \|\sigma^{Y,2}_t\|^2 \right) }{2}      +   \frac{\|\theta^1_t\|^2+\|\theta_t^2\|^2}{2}+\frac{\gamma\|\lambda_t-\gamma\sigma^{Y,1}_t+\theta^1_t\|^2}{2(1-\gamma)} \bigg)dt \\
        & \ +   \left( \gamma \tilde{\xi}_t + \theta^1_t \right)^\top d{\bf B}^1_t + \left( \theta^2_t - \gamma\sigma^{Y,2}_t \right)^\top d{\bf B}^2_t \\
        = &  \  -\frac{\gamma(1-\gamma)}{2}\left\| \tilde{\xi}_t - \frac{\lambda_t-\sigma^{Y,1}_t + \theta^1_t}{1-\gamma} \right\|^2  dt +  \left( \gamma \tilde{\xi}_t + \theta^1_t \right)^\top d{\bf B}^1_t + \left( \theta^2_t - \gamma\sigma^{Y,2}_t \right)^\top d{\bf B}^2_t ,
    \end{align*}
in which the last equality is due to \eqref{eq:V:power:2}. Hence, since $\pi\in\mathcal{A}$, for any $0\leq s\leq t$,
    \begin{equation}
    \label{eq:U:power:supermartingale}
        U(X_t^\pi,t) = U(X_s^\pi,s)\exp\left(-\int_s^t  \frac{\gamma(1-\gamma)}{2}\left\| \tilde{\xi}_l - \frac{\lambda_l-\sigma^{Y,1}_l + \theta^1_l}{1-\gamma} \right\|^2 dl\right)\mathcal{E}_{s,t},
    \end{equation}
where $\{\mathcal{E}_{s,t}\}_{t\geq s}$ satisfies $\mathcal{E}_{s,s}=1$ and, for any $0\leq s\leq t$,
\begin{equation}
d\mathcal{E}_{s,t} = \mathcal{E}_{s,t}\left( (\gamma\tilde{\xi}_t+\theta^1_t)^\top d{\bf B}^1_t + (\theta^2_t - \gamma \sigma^{Y,2}_t)^\top d{\bf B}^2_t \right).
\label{eq:dolean_exp}
\end{equation}

We shall show that, for any $t\geq 0$, 
    \begin{equation}
\label{eq:power:P}
\int_0^t \|\gamma\tilde{\xi}_s+\theta^1_s\|^2 ds = \int_0^t \left\| \frac{\gamma X^\pi_s(\Sigma^\top_s \pi_s-\sigma^{Y,1}_s)-X^{\hat{\pi},0}_s\beta_s }{\Tilde{X}_s} + \theta^1_s \right\|^2 ds < \infty,  \  \mathbb{P}\text{-a.s.}.
\end{equation}
This then implies that $\{\mathcal{E}_{s,t}\}_{t\geq s}$ in \eqref{eq:dolean_exp} is an $\mathbb{F}$-local martingale. Since $\pi \in \mathcal{A}$, we have $X^\pi\pi\in \mathcal{L}^2_n$. Consider the process $\bar{X}=\{\bar{X}_t\}_{t\geq 0}$, which satisfies $\bar{X}_0=x_0$ and, for any $t\geq 0$,
 \begin{align*}
d\bar{X}_t =& \left(p_t + (X^\pi_t\pi_t)^\top \Sigma_t(\lambda_t - \sigma^{Y,1}_t)\right)dt  + (X^\pi_t \pi_t)^\top\Sigma_t d{\bf B}^1_t   \\ &\ +\bar{X}_t \left( \left(   - \mu^Y_t + \|\sigma^{Y,1}_t\|^2 + \|\sigma^{Y,2}_t\|^2 \right)  dt     -(\sigma^{Y,1}_t)^\top d{\bf B}^1_t - (\sigma^{Y,2}_t)^\top d{\bf B}^2_t \right).
\end{align*}
Using the fact that $X^\pi\pi\in \mathcal{L}^2_n$ and Lemma \ref{lemma:sde}, we infer that $\bar{X}\in \mathcal{L}^2_1$. Notice that the difference $X^{\pi}-\bar{X}$ satisfies $X^{\pi}_0-\bar{X}_0 = 0$, and, for any $t\geq 0$, 
 \begin{equation}
 \label{eq:xbar-x}
d(\bar{X}_t-X^\pi_t) = (\bar{X}_t-X^\pi_t) \left( \left(   - \mu^Y_t + \|\sigma^{Y,1}_t\|^2 + \|\sigma^{Y,2}_t\|^2 \right)  dt     -(\sigma^{Y,1}_t)^\top d{\bf B}^1_t - (\sigma^{Y,2}_t)^\top d{\bf B}^2_t \right).
\end{equation}
By the uniqueness of the solution to \eqref{eq:xbar-x}, we infer that $\bar{X}$ and $X^\pi$ are indistinguishable, and hence $X^\pi \in \mathcal{L}^2_1$. Also, recall that $X^{\hat{\pi},0}\in \mathcal{L}^2_1$ (see the proof of Proposition \ref{prop:non-empty}). Finally, also by Lemma \ref{lemma:sde}, $X^\pi$ and $X^{\hat{\pi},0}$ admit continuous version, and thus $\Tilde{X}$ also admits continuous sample paths $\mathbb{P}$-a.s.. Together with the admissibility condition $\tilde{X}_t=X^\pi_t-X^{\hat{\pi},0}_t>0$ for a.a. $(t,\omega)\in \mathbb{R}_{+}\times \Omega$, we have $\sup_{s\in [0,t]}|\tilde{X}_s|^{-1}<\infty$, $\mathbb{P}$-a.s., for any $t\geq 0$. Therefore, these together imply that
\begin{equation*}
\begin{aligned}
& \int_0^t \left\| \frac{\gamma X^\pi_s(\Sigma^\top_s \pi_s-\sigma^{Y,1}_s)-X^{\hat{\pi},0}_s\beta_s }{\Tilde{X}_s} + \theta^1_s \right\|^2 ds \\
\leq &\ 2\int_0^t \left(2\gamma^2\left(\sup_{s\in[0,t]}|\Tilde{X}_s|^{-1}\right)^2\left( 2\left( \|\Sigma_s\|^2\|X^\pi_s\pi\|^2 + \|\sigma^{Y,1}_s\|^2|X^\pi_s|^2 \right) + \|\beta_s\|^2|X^{\hat{\pi},0}_s|^2 \right) + \|\theta^1_s\|^2\right) ds \\
<&\ \infty , \  \mathbb{P}\text{-a.s.,}
\end{aligned}
\end{equation*}
which proves \eqref{eq:power:P}. 
Since $\{\mathcal{E}_{s,t}\}_{t\geq s}$ in \eqref{eq:dolean_exp} is bounded below by zero, it is also an $\mathbb{F}$-super-martingale. 
With $0<\gamma<1$, by \eqref{eq:U:power:supermartingale}, for any $0\leq s\leq t$,
\begin{equation*}
\mathbb{E}[U(X^\pi_t,t)|\mathcal{F}_s] \leq  U(X^\pi_s,s) \mathbb{E}[\mathcal{E}_{s,t}|\mathcal{F}_s] \leq U(X^\pi_s,s),
\end{equation*}
and hence $\{U(X^\pi_t,t)\}_{t\geq 0}$ is also an $\mathbb{F}$-super-martingale.\\

In particular, with $\pi\in\mathcal{A}$ given by $\pi^*$, we have, for any $t\geq 0$,
\begin{equation*}
\xi^*_t := \tilde{\xi}^{\pi^*}_t = \frac{X^*_t(\Sigma^\top_t \pi^*_t-\sigma^{Y,1}_t)-X^{\hat{\pi},0}_t\beta_t }{X^*_t-X^{\hat{\pi},0}_t} =\frac{\lambda_t-\sigma^{Y,1}_t+\theta^1_t}{1-\gamma},
\end{equation*}
and correspondingly, from \eqref{eq:U:power:supermartingale}, for any $0\leq s\leq t$,
\begin{equation*}
U(X_t^*,t) = U(X_s^*,s) \mathcal{E}_{s,t}^*,
\end{equation*}
where  $\{\mathcal{E}_{s,t}^*\}_{t\geq s}$ satisfies $\mathcal{E}_{s,s}^*=1$ and, for any $0\leq s\leq t$, 
\begin{equation*}
d\mathcal{E}_{s,t}^* = \mathcal{E}_{s,t}^*\left(  \frac{\left(\gamma(\lambda_t - \sigma^{Y,1}_t)+  \theta_t^1  \right)^\top}{1-\gamma} d{\bf B}^1_t + \left( \theta_t^2 - \gamma \sigma^{Y,2}_t   \right)^\top d{\bf B}^2_t\right),
\end{equation*}
which is clearly an $\mathbb{F}$-martingale, by the (uniform) boundedness of $\lambda_{\cdot}$, $\sigma^{Y,1}_{\cdot}$, $\sigma^{Y,2}_{\cdot}$, $\theta^1$, and $\theta^2$. Therefore, $\{U(X_t^*,t)\}_{t\geq 0}$ is also an $\mathbb{F}$-martingale.\\

Finally, by It\^o's lemma on the translated random field $\Tilde{U}$, for any $\tilde{x}\in\mathbb{R}_{++}$ and $t\geq 0$, 
    \begin{equation*}
        d\tilde{U}(\tilde{x},t) = \tilde{U}(\tilde{x},t) \left(  \left( v_t + \frac{\|\theta^1_t\|^2+\|\theta^2_t\|^2}{2}\right)dt + (\theta^1_t)^\top d{\bf B}^1_t +  (\theta^2_t)^\top d{\bf B}^2_t \right),
    \end{equation*}
which coincides with \eqref{eq:SPDE}, with $\nu_t = p_t + X^{\hat{\pi},0}_t\alpha_t$, $\kappa^1_t = X^{\hat{\pi},0}_t\beta_t$, and $\kappa^2_t = -X^{\hat{\pi},0}_t\sigma^{Y,2}_t$, for any $t\geq 0$, as well as $a_1$ and $a_2$ given in \eqref{eq:vol:power}.

    \section{Proof of Theorem \ref{pp:exp}}

    \label{sec:app:proof:exp}

As this proof follows similarly to that of Theorem \ref{pp:power}, only the essential steps are outlined. First, $\pi^*$ can be rewritten as, for any $t\geq 0$,
\begin{equation*}
\pi_t^* = \hat{\pi}_t+\frac{\Gamma_t}{X^*_t}\left(\left(\Sigma^{\top}_t\right)^{-1}\left(\lambda_t+\theta^1_t\right)-\hat{\pi}_t\right),
\end{equation*}
and thus $\pi^*\in\tilde{\mathcal{A}}=\mathcal{A}$, with $\xi\equiv\left(\Sigma^{\top}_\cdot\right)^{-1}\left(\lambda_\cdot+\theta^1\right)-\hat{\pi}\in\mathcal{L}^{2}_{n,\text{BMO}}$, by the (uniform) boundedness and Proposition \ref{pp:admissible:exp}. Next, since $\Gamma$ is uniformly positive, $U$ in \eqref{eq:U:exp} satisfies (i) and (ii) in Definition \ref{def:forward}. To verify (iii) of Definition \ref{def:forward}, it suffices to show that $\{U(X^\pi_t,t)\}_{t\geq 0}$ is an $\mathbb{F}$-super-martingale for any $\pi\in\mathcal{A}$,  and that $\{U(X^*_t,t)\}_{t\geq 0}$ is an $\mathbb{F}$-martingale.\\

For any $\pi\in\mathcal{A}$, define $R^{\pi}_t:=-\exp\left( - \frac{X^{\pi}_t-X^{\hat{\pi},0}_t}{\Gamma_t} + V_t \right)$ for $t\geq 0$. By It\^o's lemma, for any $t\geq 0$,
\begin{align*}
\frac{dR^{\pi}_t}{R^{\pi}_t}=&\;\frac{1}{2}\left(\frac{X^{\pi}_t}{\Gamma_t}\right)^2\left\|\Sigma^\top_t\pi_t - \sigma^{Y,1}_t- \beta_t  -  \frac{\Gamma_t}{X^{\pi}_t}\left(\lambda_t-\sigma^{Y,1}_t+\theta^1_t - \beta_t\right) \right\|^2 dt \\
        & -\left(\frac{X^{\pi}_t}{\Gamma_t} (\Sigma^\top_t\pi_t-\sigma^{Y,1}_t -\beta_t ) -\theta^1_t \right)^{\top}  d{\bf B}^1_t + \left(\theta^2_t\right)^\top d{\bf B}^2_t.
\end{align*}
Hence, for any $0\leq s\leq t$,
\begin{align*}
U\left(X^{\pi}_t,t\right)=&\;U\left(X^{\pi}_s,s\right)\\&\times\exp\left(\int_{s}^{t}\frac{1}{2}\left(\frac{X^{\pi}_l}{\Gamma_l}\right)^2\left\|\Sigma^\top_l\pi_l - \sigma^{Y,1}_l- \beta_l  -  \frac{\Gamma_l}{X^{\pi}_l}\left(\lambda_l-\sigma^{Y,1}_l+\theta^1_l - \beta_l\right) \right\|^2dl\right)\mathcal{E}_{s,t},
\end{align*}
where $\left\{\mathcal{E}_{s,t}\right\}_{t\geq s}$ satisfies $\mathcal{E}_{s,s}=1$ and, for any $0\leq s\leq t$,
\begin{equation*}
d\mathcal{E}_{s,t}=\mathcal{E}_{s,t}\left(-\left(\frac{X^{\pi}_t}{\Gamma_t} (\Sigma^\top_t\pi_t-\sigma^{Y,1}_t -\beta_t ) -\theta^1_t \right)^{\top}  d{\bf B}^1_t + \left(\theta^2_t\right)^\top d{\bf B}^2_t\right).
\end{equation*}
Since $\pi\in\mathcal{A}=\tilde{\mathcal{A}}$, $\frac{X^{\pi}}{\Gamma} (\Sigma^\top_\cdot\pi-\sigma^{Y,1}_\cdot -\beta )\equiv\Sigma_\cdot^\top\xi\in\mathcal{L}^{2}_{n,\text{BMO}}$, and hence, together with the uniform boundedness, the process $\left\{\mathcal{E}_{s,t}\right\}_{t\geq s}$ is an $\mathbb{F}$-martingale.\\

Therefore, for any $\pi\in\mathcal{A}$, and for any $0\leq s\leq t$,
\begin{align*}
&\;\mathbb{E}\left[\exp\left(\int_{s}^{t}\frac{1}{2}\left(\frac{X^{\pi}_l}{\Gamma_l}\right)^2\left\|\Sigma^\top_l\pi_l - \sigma^{Y,1}_l- \beta_l  -  \frac{\Gamma_l}{X^{\pi}_l}\left(\lambda_l-\sigma^{Y,1}_l+\theta^1_l - \beta_l\right) \right\|^2dl\right)\mathcal{E}_{s,t}\Big\vert\mathcal{F}_s\right]\\\geq&\;\mathbb{E}\left[\mathcal{E}_{s,t}\vert\mathcal{F}_s\right]=\mathcal{E}_{s,s}=1,
\end{align*}
where the equality holds when $\pi\equiv\pi^*$. By the uniform negativity of $U$ in \eqref{eq:U:exp}, for any $\pi\in\mathcal{A}$, and for any $0\leq s\leq t$,
\begin{equation*}
\mathbb{E}\left[U\left(X^{\pi}_t,t\right)\vert\mathcal{F}_s\right]\leq U\left(X^{\pi}_s,s\right),
\end{equation*}
and in particular, $\mathbb{E}\left[U\left(X^*_t,t\right)\vert\mathcal{F}_s\right]=U\left(X^*_s,s\right)$. These show that $\{U(X^\pi_t,t)\}_{t\geq 0}$ is an $\mathbb{F}$-super-martingale for any $\pi\in\mathcal{A}$,  and that $\{U(X^*_t,t)\}_{t\geq 0}$ is an $\mathbb{F}$-martingale.\\

Finally, by It\^o's lemma on $U$ in \eqref{eq:U:exp}, for any $x\in\mathbb{R}$ and $t\geq 0$,
\begin{align*}
dU\left(x,t\right)=&\;U\left(x,t\right)\left(\left(\frac{1}{2}\|\lambda_t-\sigma^{Y,1}_t+\theta^1_t-\beta_t\|^2+\frac{p_t}{\Gamma_t}+x\left(\beta_t^\top\theta^1_t-\left(\sigma^{Y,2}_t\right)^\top\theta^2_t\right)\right.\right.\\&\;\left.\left.\quad\quad\quad\quad\;\;+\frac{x}{\Gamma_t}\left(\alpha_t-\|\beta_t\|^2-\|\sigma^{Y,2}_t\|^2\right)+\frac{1}{2}\left(\frac{x}{\Gamma_t}\right)^2\left(\|\beta_t\|^2+\|\sigma^{Y,2}_t\|^2\right)\right)dt\right.\\&\;\left.\quad\quad\quad\quad+\left(\theta^1_t+\frac{x}{\Gamma_t}\beta_t\right)^\top d{\bf B}^1_t+\left(\theta^2_t-\frac{x}{\Gamma_t} \sigma^{Y,2}_t\right)^\top d{\bf B}^2_t\right),
\end{align*}
which shows that the volatility processes of $U$ in \eqref{eq:U:exp} are indeed given by \eqref{eq:vol:exp}.

\end{appendices}}

\end{document}